\documentclass{IEEEtran}
\usepackage{cite}
\usepackage{amsmath,amssymb,amsfonts}
\usepackage{algorithmic}
\usepackage{graphicx,subfigure}
\usepackage{textcomp}
\def\BibTeX{{\rm B\kern-.05em{\sc i\kern-.025em b}\kern-.08em
    T\kern-.1667em\lower.7ex\hbox{E}\kern-.125emX}}
\begin{document}
\title{Machine-Learning Classification of Closed and Open Radiating Wires from Near Magnetic or Electric Field Scan Images}
\author{Amir Geranmayeh  
\thanks{The author was with Continental Automotive GmbH, VDO-Str. 1, 64832 Babenhausen, Germany (e-mail: a.geranmayeh@gmail.com).}}

\maketitle

\begin{abstract}
Sets of intelligent classifiers are applied to the near-field scan-data in order to automatically classify the shape of radiating wirings. The support vector machine, {\it k}-nearest neighbors algorithm, and Gaussian process classifications are trained using the near-field radiation pattern of diverse radiating wire configurations. Leave-one-out cross-validation is used for estimating the performance of the predictive models. 

\end{abstract}


\section{Problem Statement} \label{sec:introduction}

\IEEEPARstart{T}{he} multilayer printed circuit boards (PCB) stack-ups normally have internal ground (GND) or power planes beneath the outer layers on which the high-speed data links or switching regulator circuitry cause radiating emissions. The electronic hardware developers commonly use two dimensional (2D) near-field scanning with the electric or magnetic probes to localize the source of radiating noises on PCB \cite{b3}. The electric or magnetic field scan images are intrinsically contributions of two elementary radiator shapes in different orientations, i.e. either they have been radiated mainly by open electric dipoles or they are predominantly coming from closed magnetic loop antennas. For electromagnetic compatibility (EMC) engineers is of great importance to detect the nature of excited modes, i.e. whether the emissions have been originally raised from electric type antenna or magnetic loop-form radiator. These two types of root cause for unintentional radiations demand different subsiding countermeasures to tackle with. These two distinct categories are equivalent to capacitive and inductive couplings in the quasi-static regime. In other words, the radiated emissions can be raised electrically from the movement of charges along the length of a conductor line or magnetically from the oscillating flux coupled through the surface of a closed conducting path. The pattern of hotspots in near-field scans are of course changed according to the electric size of the radiating objects. Although radiating source footprints change dynamically, they remain related to the geometric topology owing to the physics of electromagnetic wave propagation. Using a decoupling capacitor to shorten the return path of the noise to the supply, for instance, may turn to an open-end radiator at higher frequency bands due to the growth of the parasitic impedances.

This work investigates the possibility of classification of open and closed radiating wires in the vicinity of GND plane and the absence of any other scatterers based on only electric or magnetic near-field 2D scan data. The ultimate goal is here to find out from the captured sample images the contributation of which fundamental case has had a dominant effect, the charge accumulations in a stub endings or fluxes flowing out of a coil. Such a distinction between the configurations of emitting traces provides simplified representative model for the radiating PCB, which can in turn readily replace complex components for system level evaluations. More importantly, this type of diagnosis helps the layout designer to come up with suitable remedies for de-noising purposes. By looking at the magnitude of the field components on the scan plane, however, the human eyes may not distinguish whether the resulting ring trajectory of the secondary electric or magnetic currents have a closed or open origin. Typical sorts of unwanted electrically radiating conductors are trace routings with high terminating mismatched impedances or not well-shielded cables with imperfect pin connections, whereas the usual magnetically coupled radiating loops are common-mode currents which close their return path through the GND planes or rotational flows of differential signaling on distanced-pair bus systems.

\section{Training Dataset Construction} \label{sec:dataset}

\begin{figure}[!t] 
   \begin{center}
   \subfigure{\includegraphics[width=0.23\columnwidth]{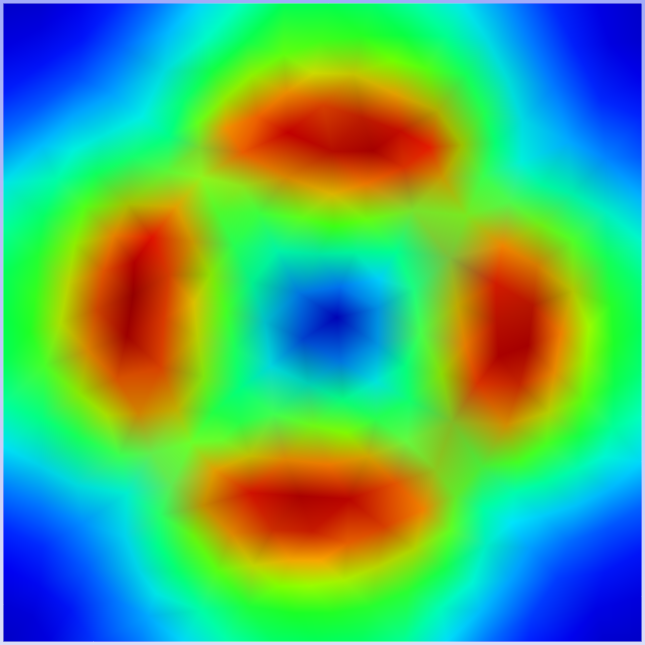}}
   \subfigure{\includegraphics[width=0.23\columnwidth]{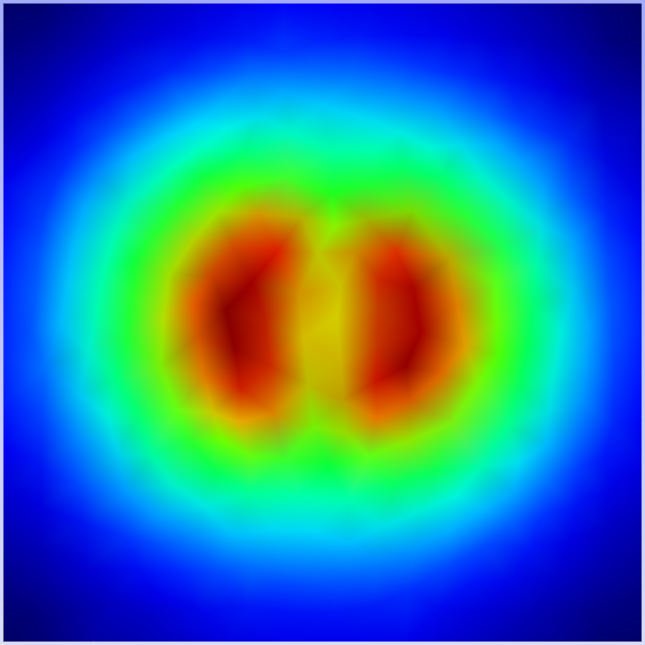}}
   \subfigure{\includegraphics[width=0.23\columnwidth]{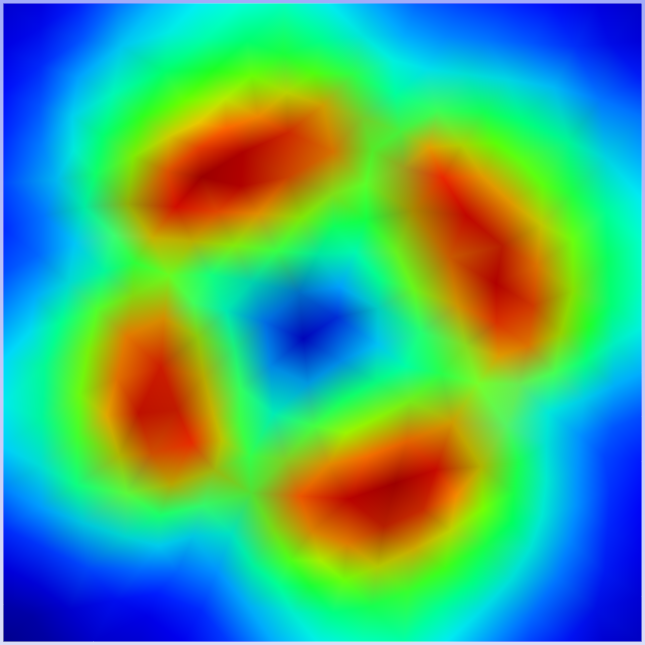}}
   \subfigure{\includegraphics[width=0.23\columnwidth]{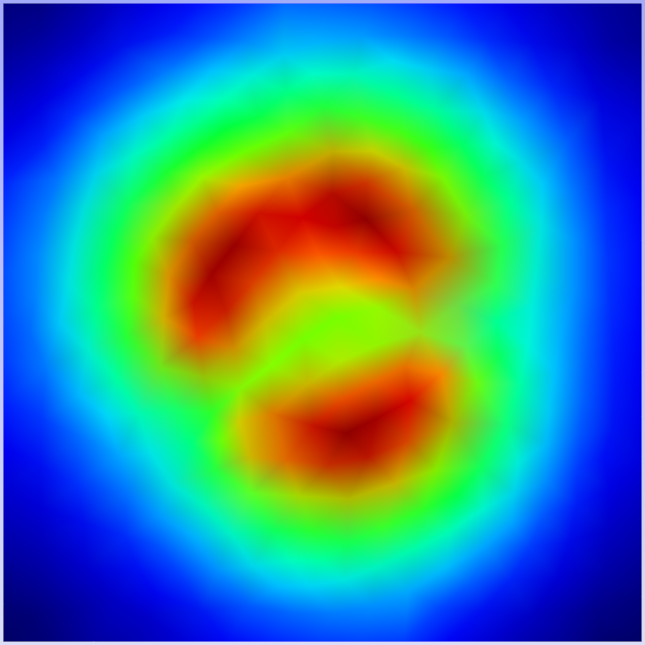}}
   \subfigure{\includegraphics[width=0.23\columnwidth]{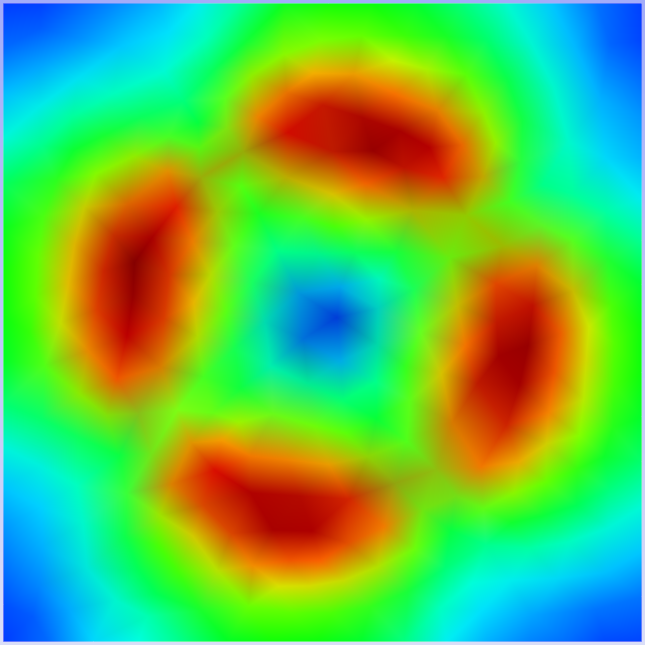}}
   \subfigure{\includegraphics[width=0.23\columnwidth]{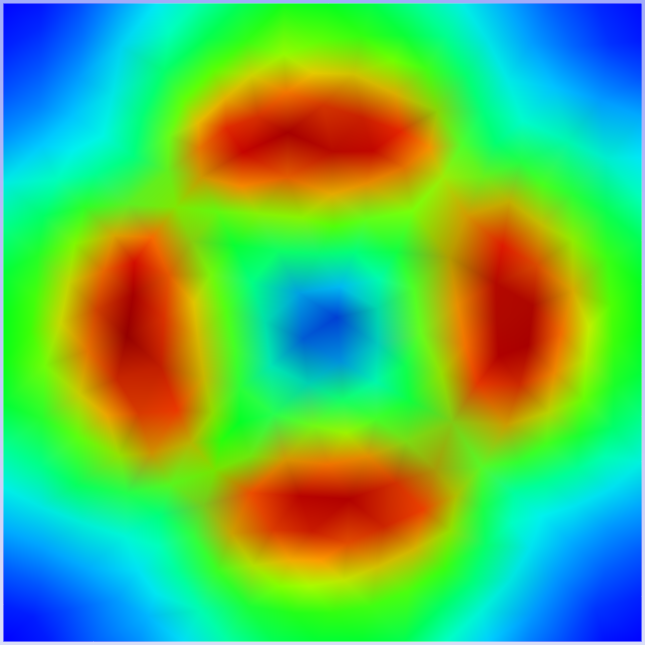}}
   \subfigure{\includegraphics[width=0.23\columnwidth]{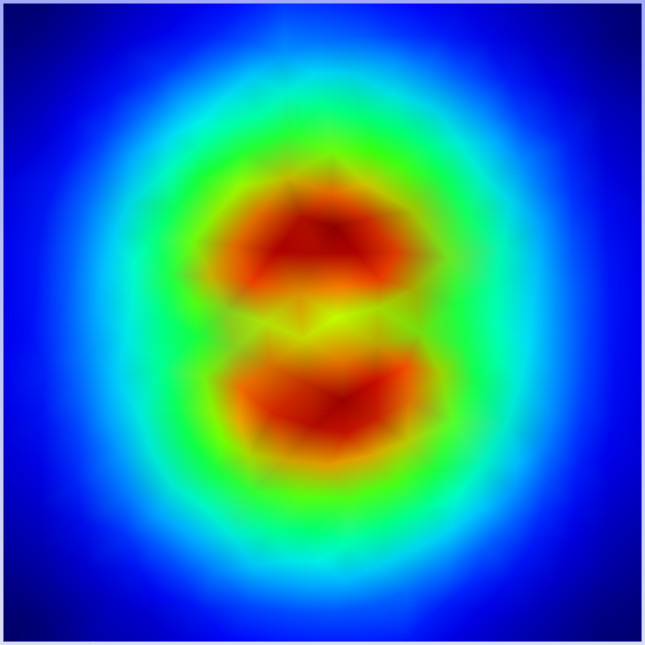}}
   \subfigure{\includegraphics[width=0.23\columnwidth]{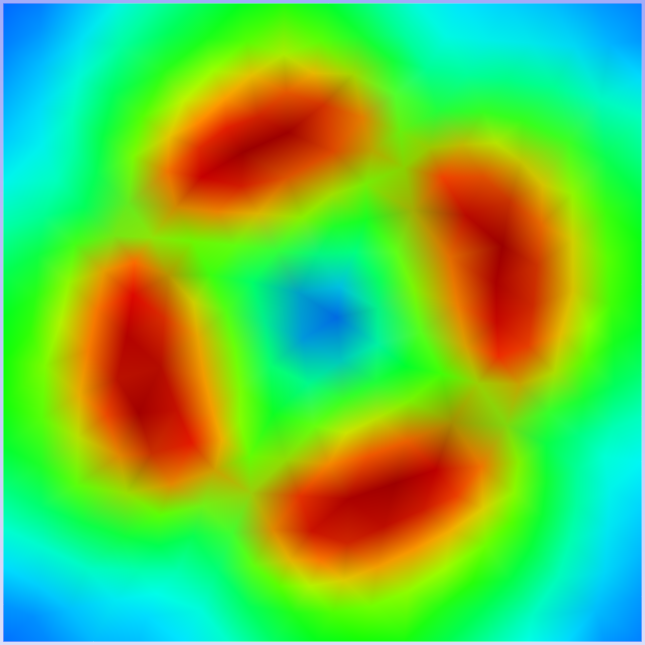}}
   \subfigure{\includegraphics[width=0.23\columnwidth]{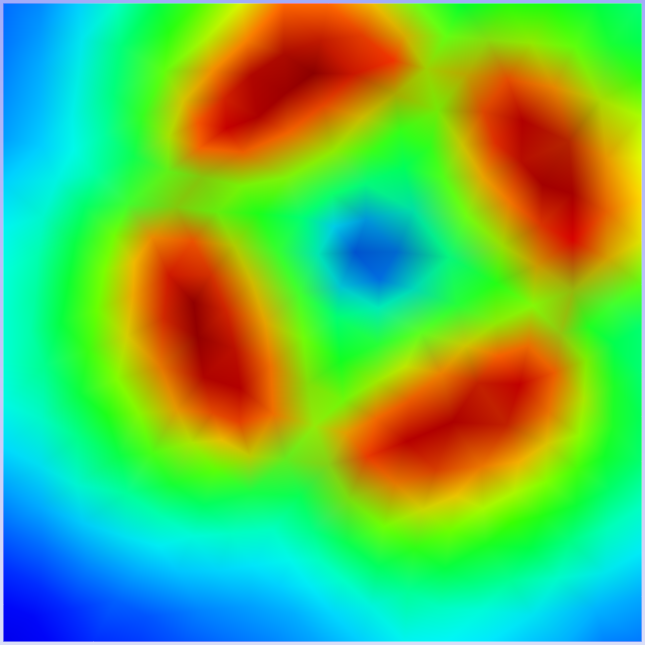}}
   \subfigure{\includegraphics[width=0.23\columnwidth]{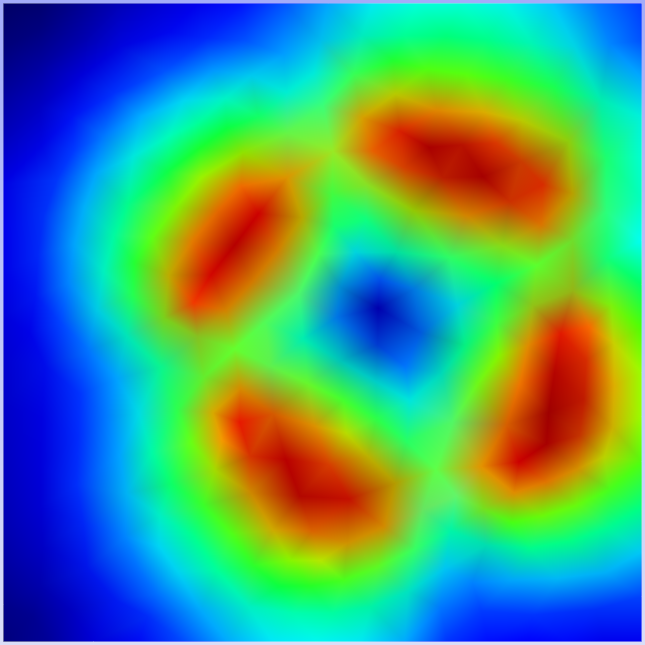}}
   \subfigure{\includegraphics[width=0.23\columnwidth]{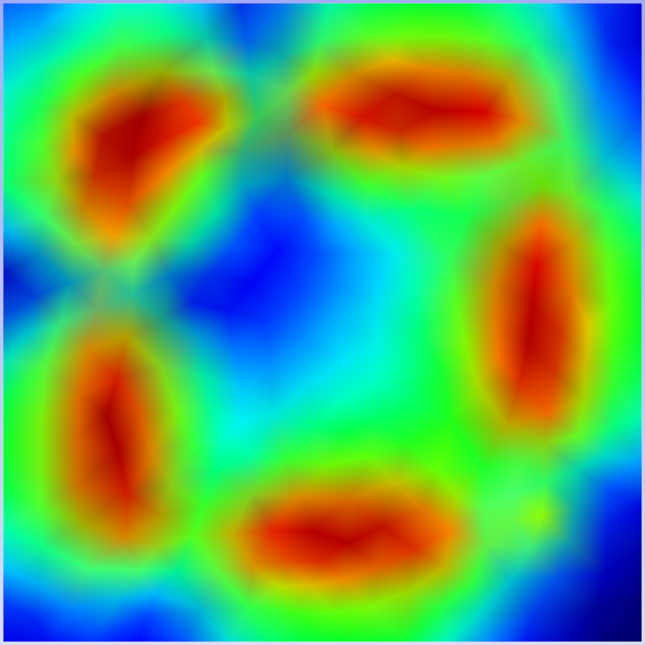}}
   \subfigure{\includegraphics[width=0.23\columnwidth]{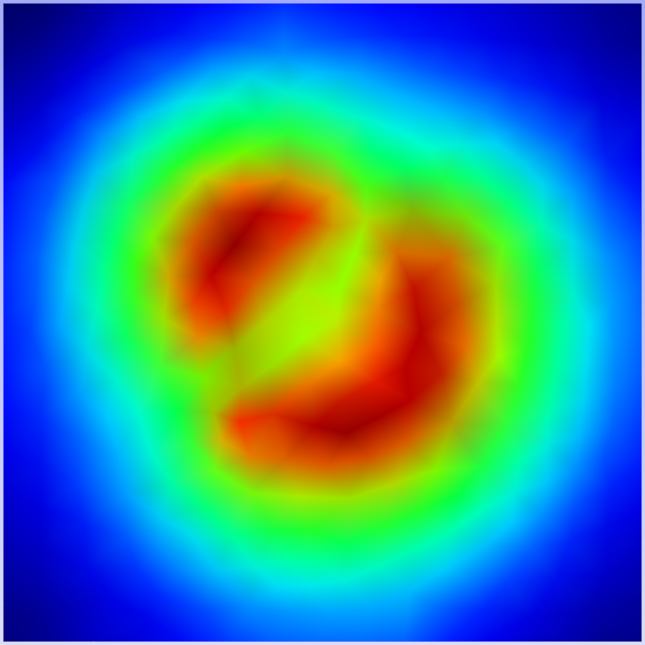}}
   \subfigure{\includegraphics[width=0.23\columnwidth]{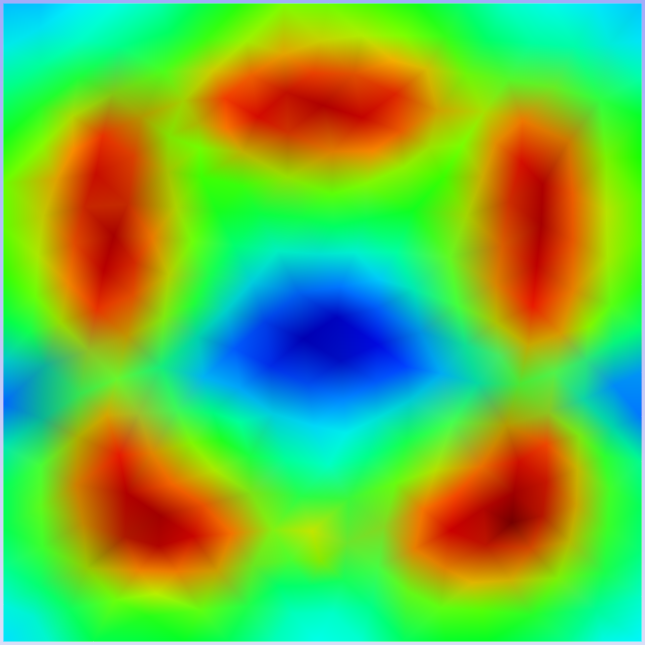}}
   \subfigure{\includegraphics[width=0.23\columnwidth]{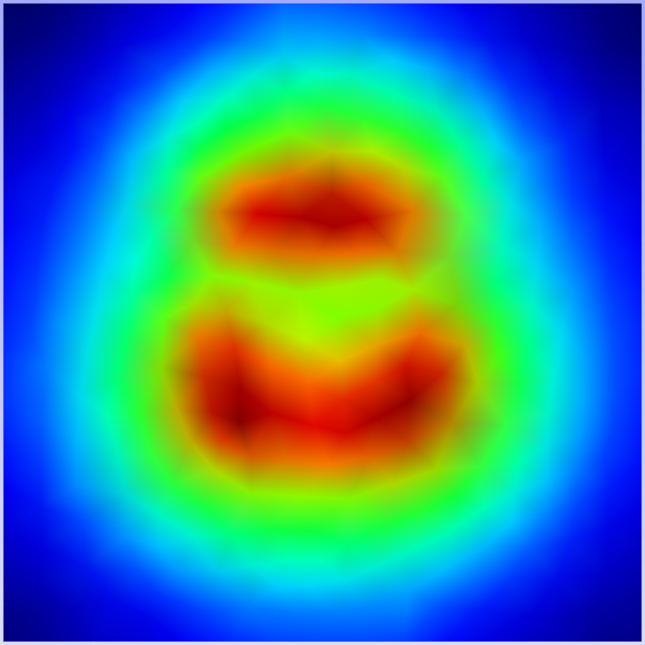}}
   \subfigure{\includegraphics[width=0.23\columnwidth]{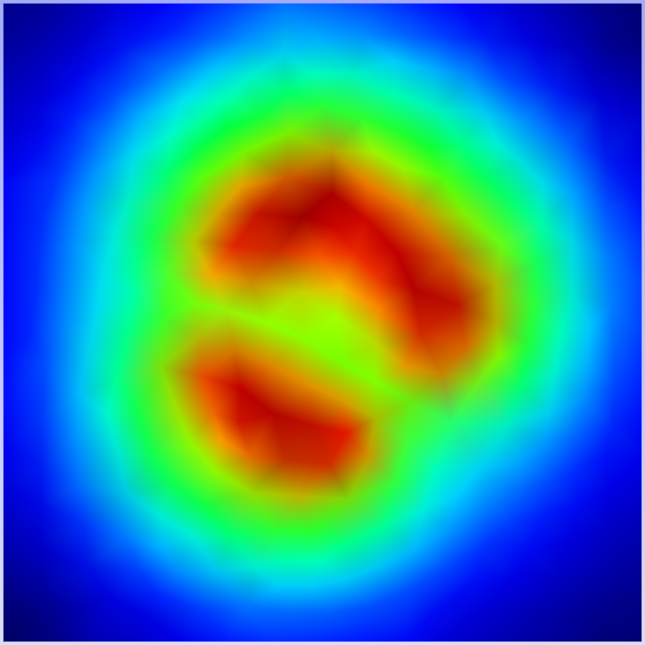}}
   \subfigure{\includegraphics[width=0.23\columnwidth]{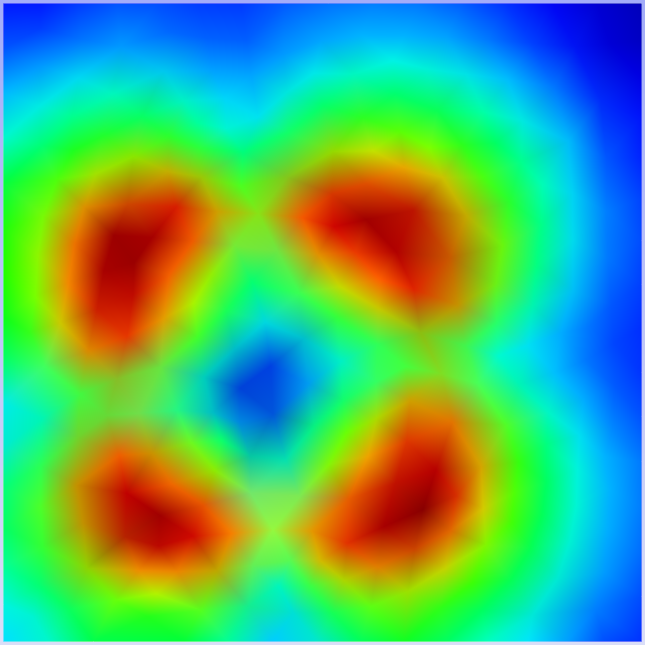}}
   \subfigure{\includegraphics[width=0.23\columnwidth]{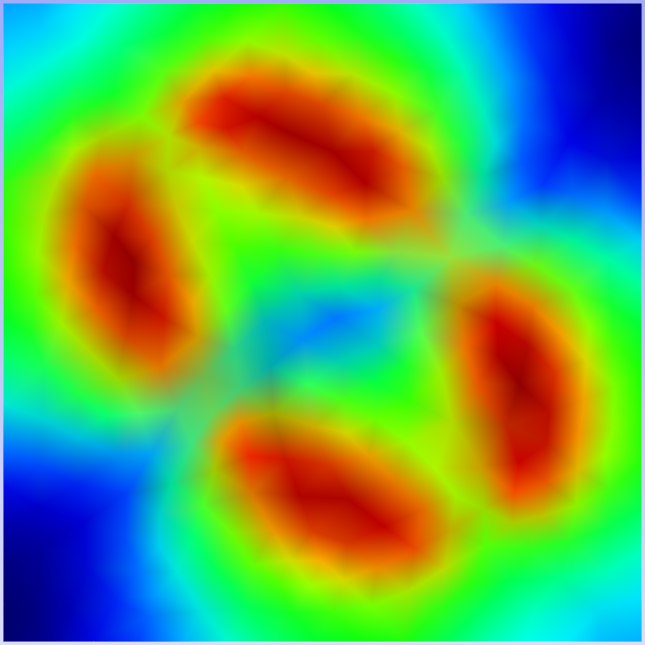}}
   \subfigure{\includegraphics[width=0.23\columnwidth]{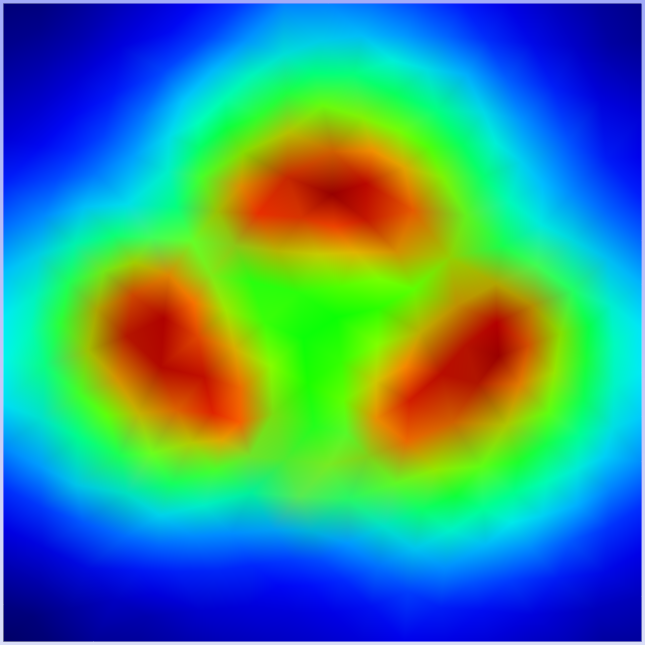}}
   \subfigure{\includegraphics[width=0.23\columnwidth]{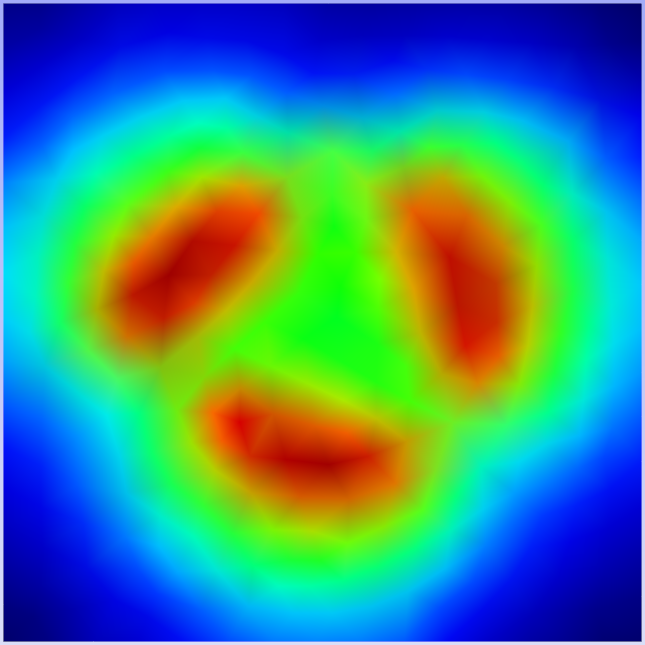}}	
   \subfigure{\includegraphics[width=0.23\columnwidth]{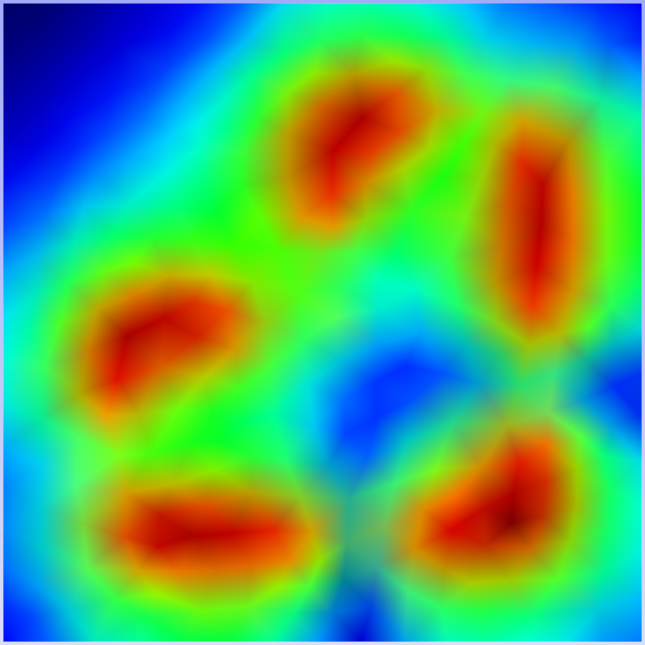}}
   \subfigure{\includegraphics[width=0.23\columnwidth]{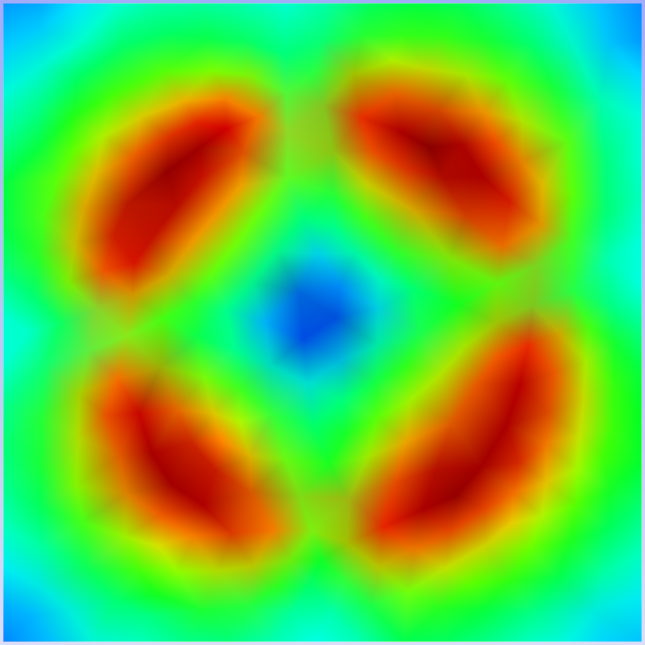}}
   \subfigure{\includegraphics[width=0.23\columnwidth]{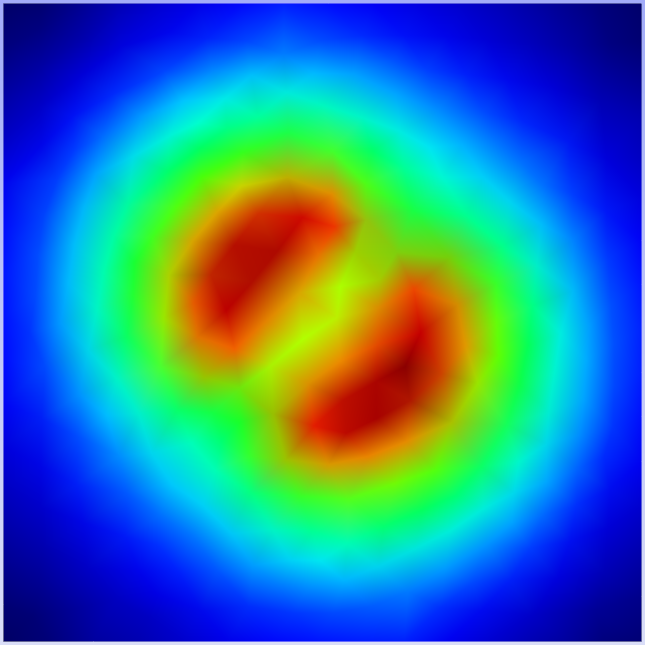}}
   \subfigure{\includegraphics[width=0.23\columnwidth]{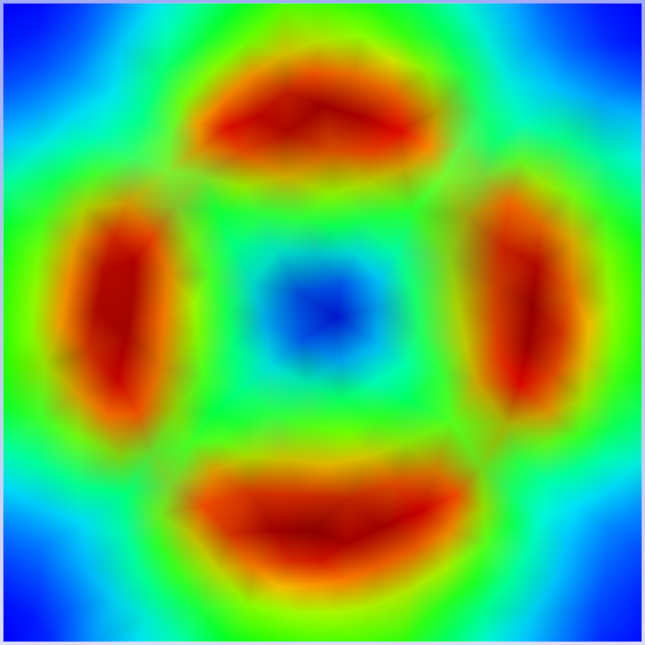}}
   \subfigure{\includegraphics[width=0.23\columnwidth]{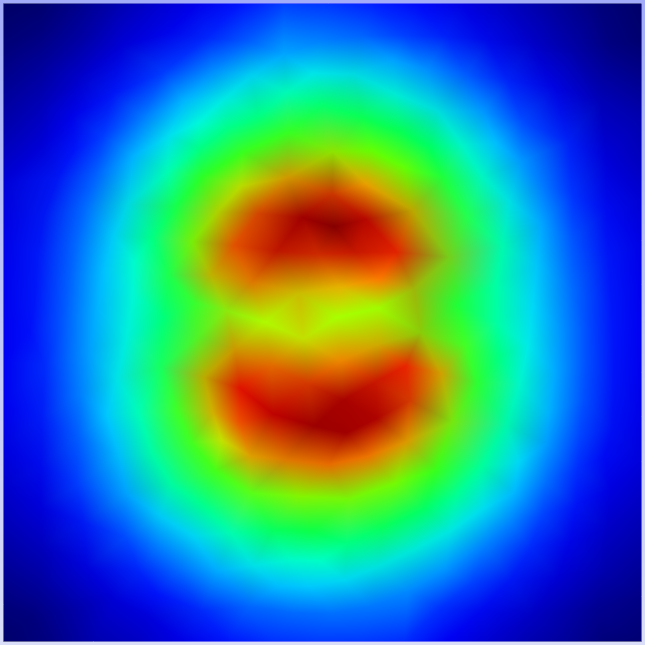}}
   \subfigure{\includegraphics[width=0.23\columnwidth]{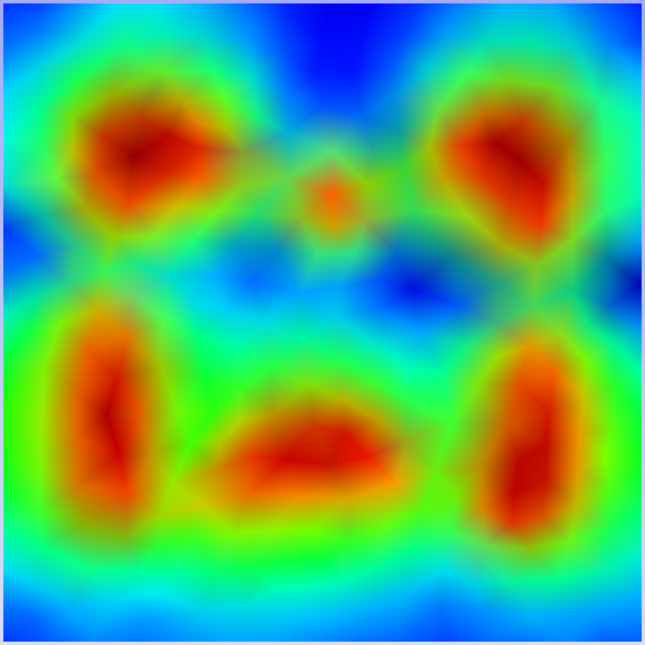}}
   \subfigure{\includegraphics[width=0.23\columnwidth]{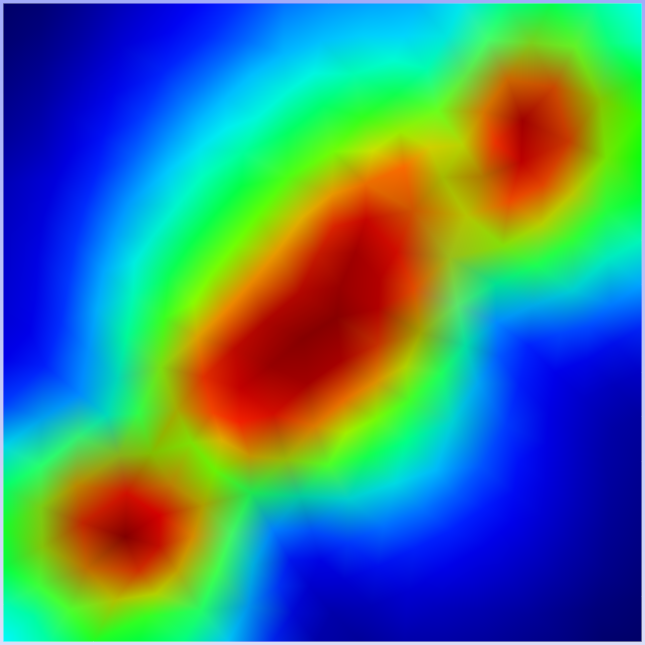}}
   \subfigure{\includegraphics[width=0.23\columnwidth]{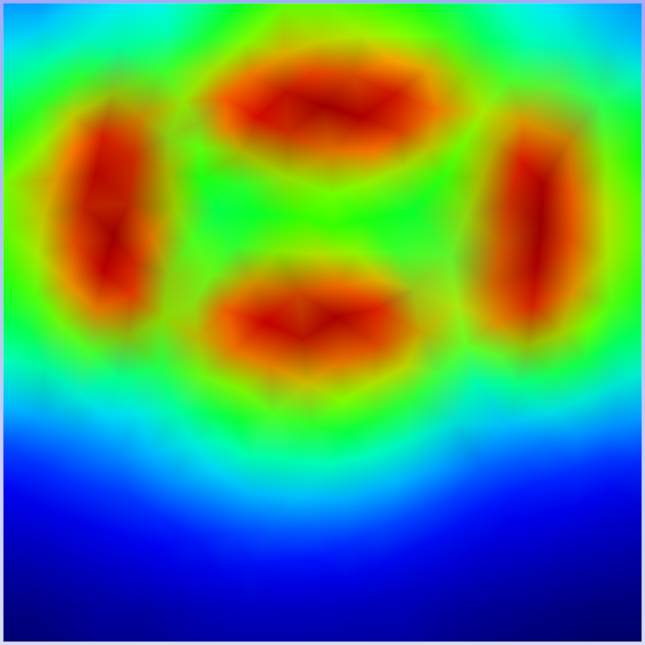}}
   \subfigure{\includegraphics[width=0.23\columnwidth]{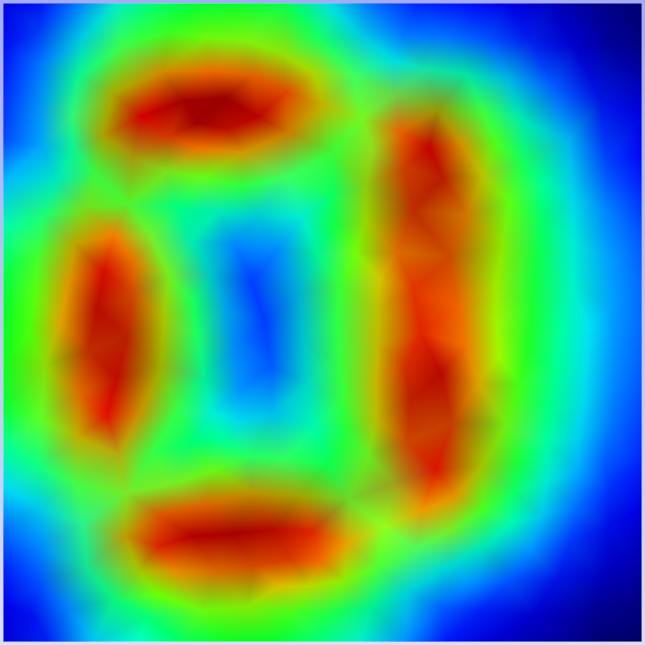}}
   \subfigure{\includegraphics[width=0.23\columnwidth]{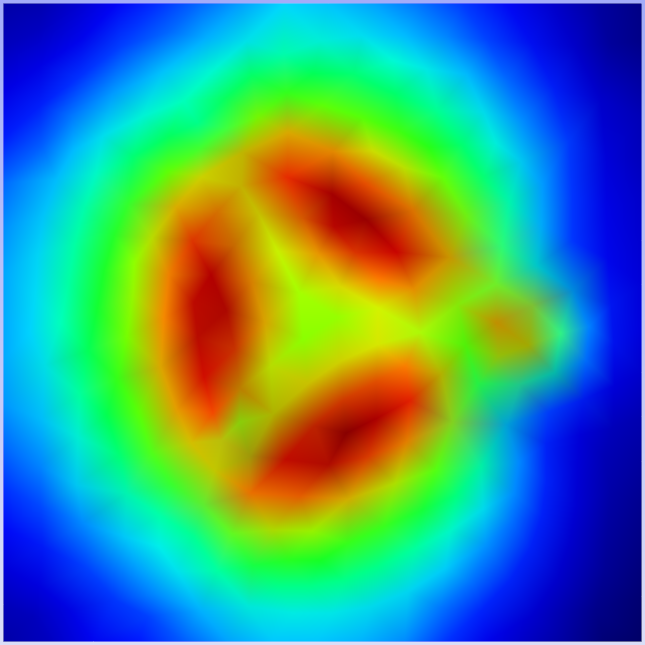}}
   \subfigure{\includegraphics[width=0.23\columnwidth]{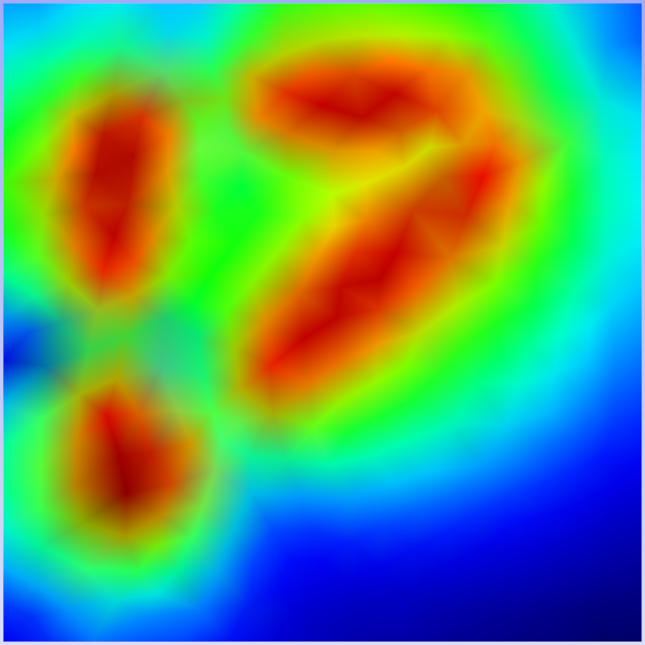}}
   \subfigure{\includegraphics[width=0.23\columnwidth]{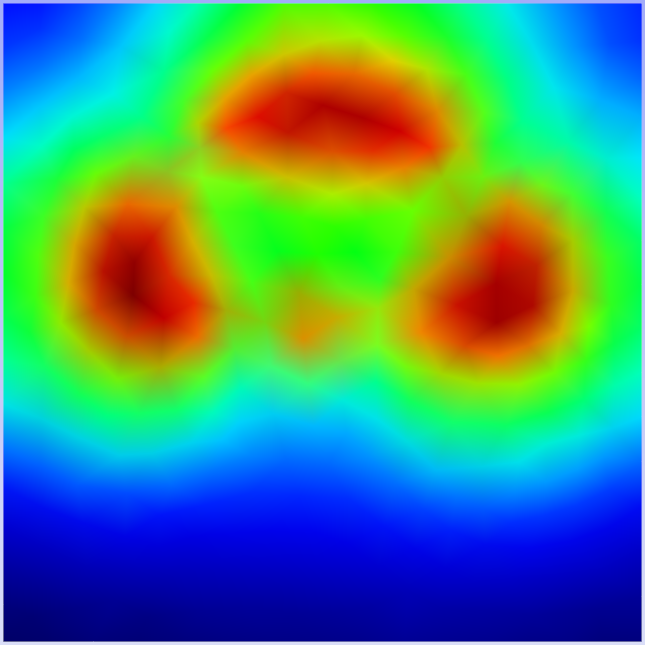}}
   \subfigure{\includegraphics[width=0.23\columnwidth]{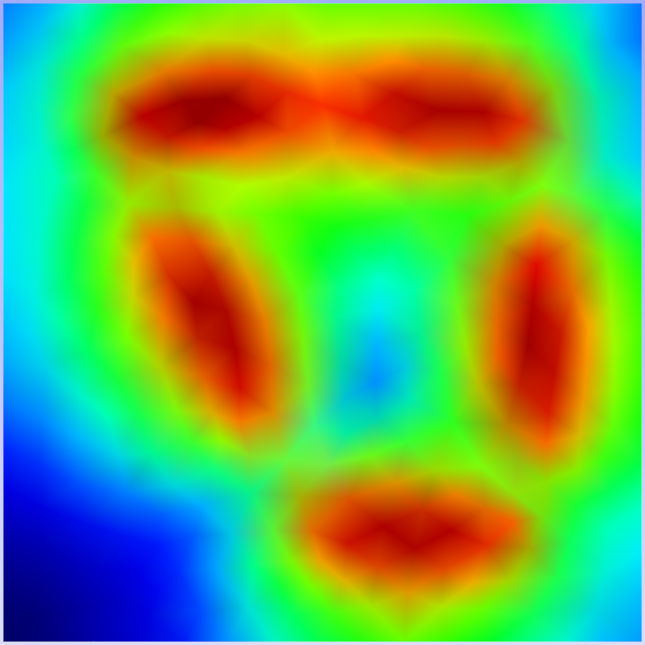}}
   \end{center} 
\caption{Magnitude of the radiated magnetic field from 32 radiating closed wires with different shapes.} 
\label{Fig1} 
\end{figure}

\begin{figure}[!t] 
   \begin{center}
   \subfigure{\includegraphics[width=0.23\columnwidth]{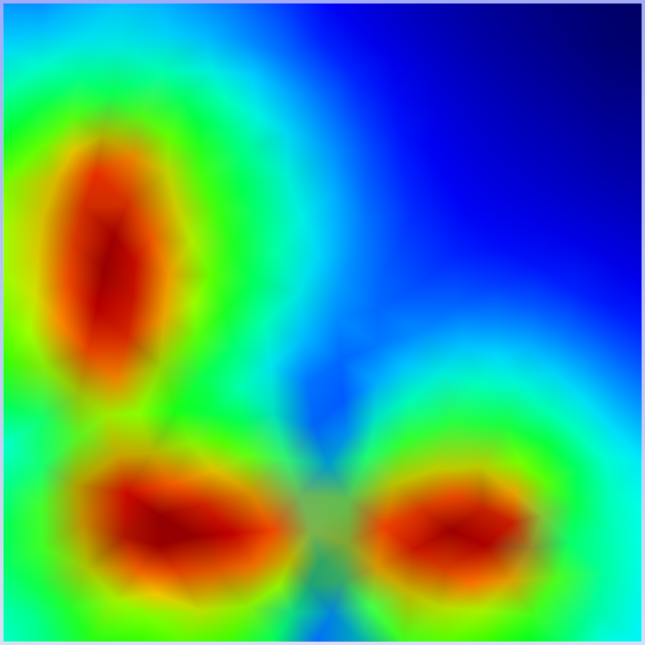}}
   \subfigure{\includegraphics[width=0.23\columnwidth]{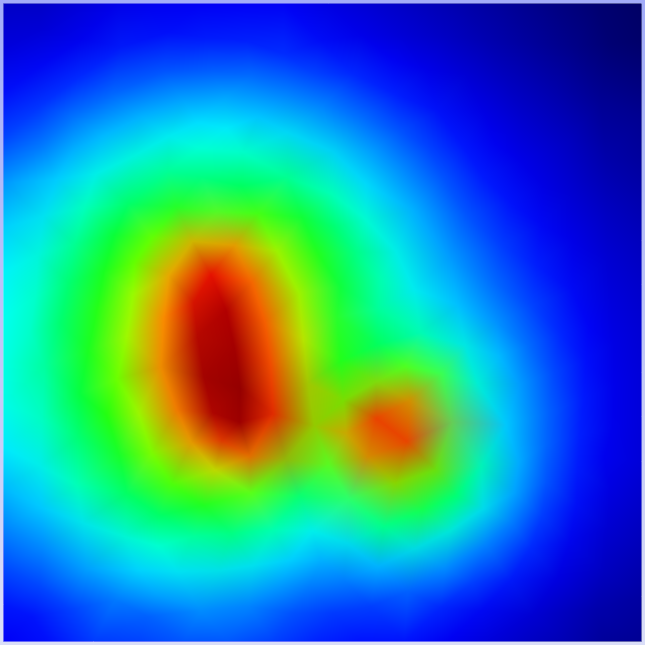}}
   \subfigure{\includegraphics[width=0.23\columnwidth]{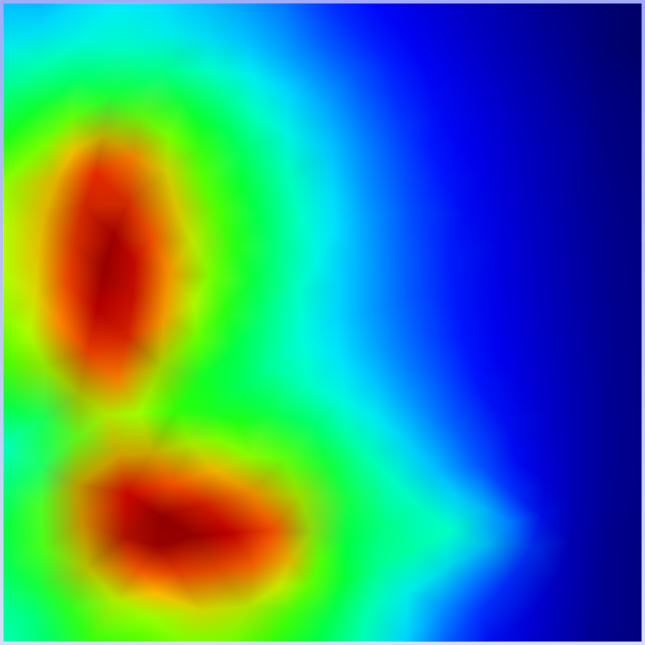}}
   \subfigure{\includegraphics[width=0.23\columnwidth]{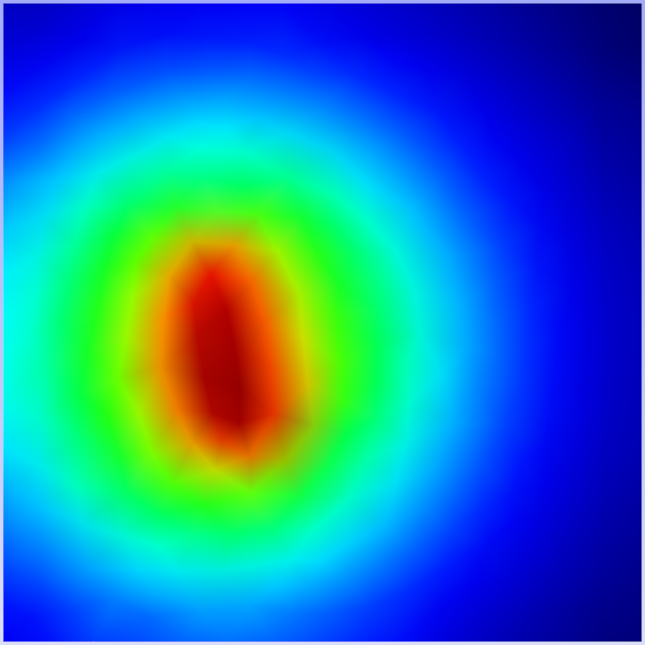}}
   \subfigure{\includegraphics[width=0.23\columnwidth]{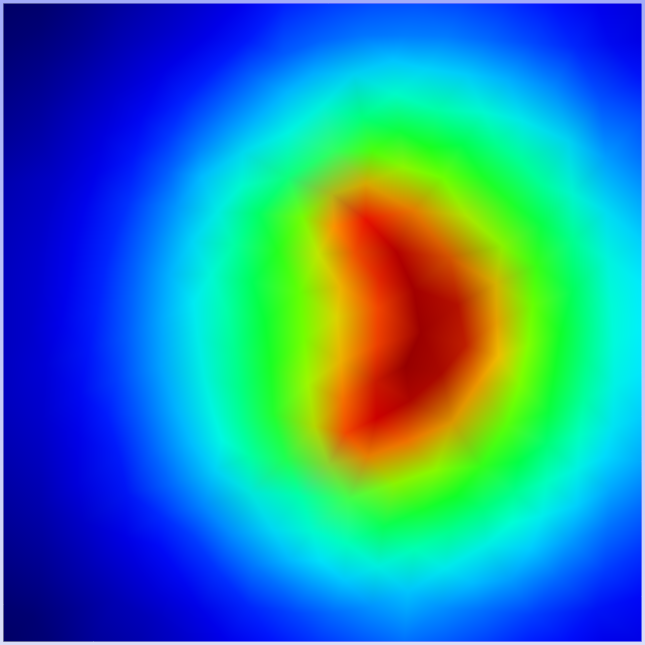}}
   \subfigure{\includegraphics[width=0.23\columnwidth]{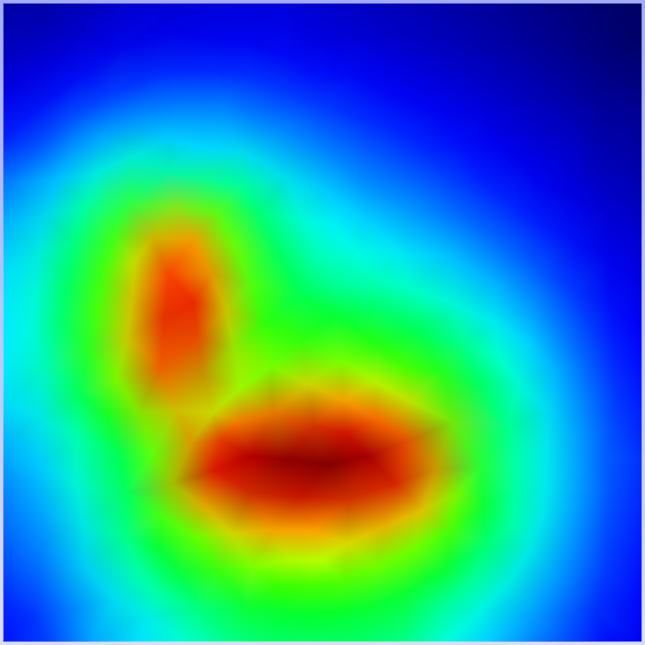}}
   \subfigure{\includegraphics[width=0.23\columnwidth]{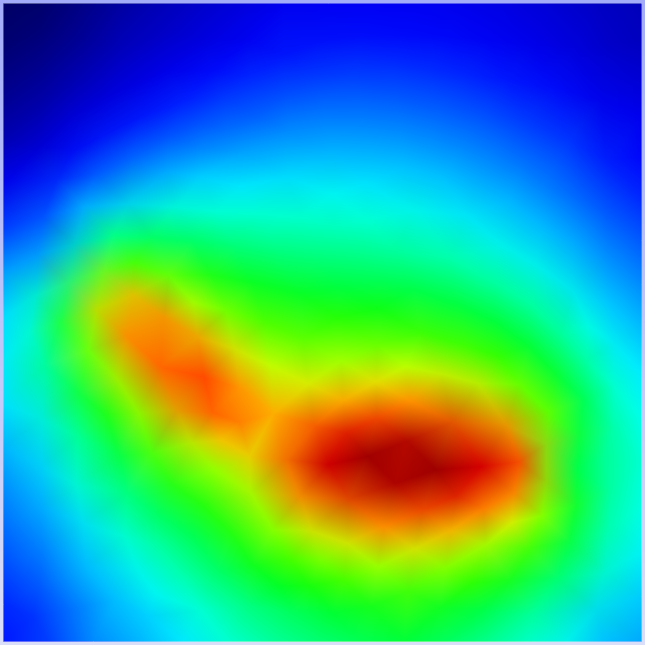}}
   \subfigure{\includegraphics[width=0.23\columnwidth]{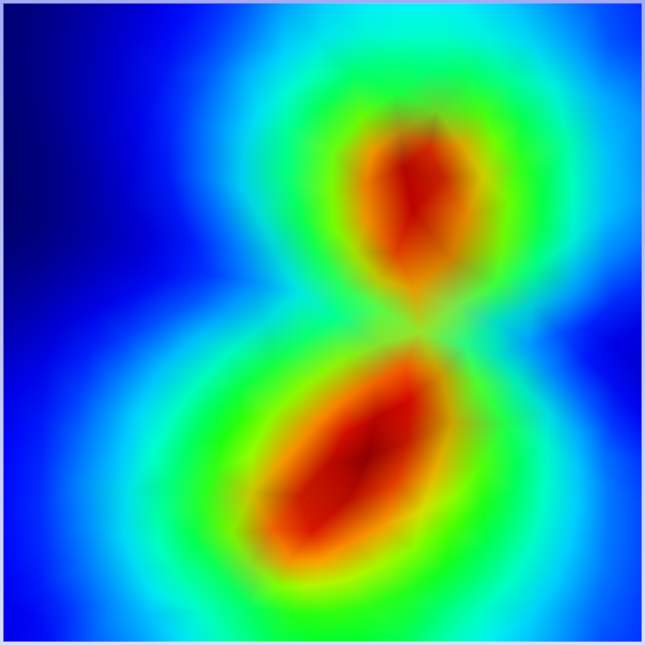}}
   \subfigure{\includegraphics[width=0.23\columnwidth]{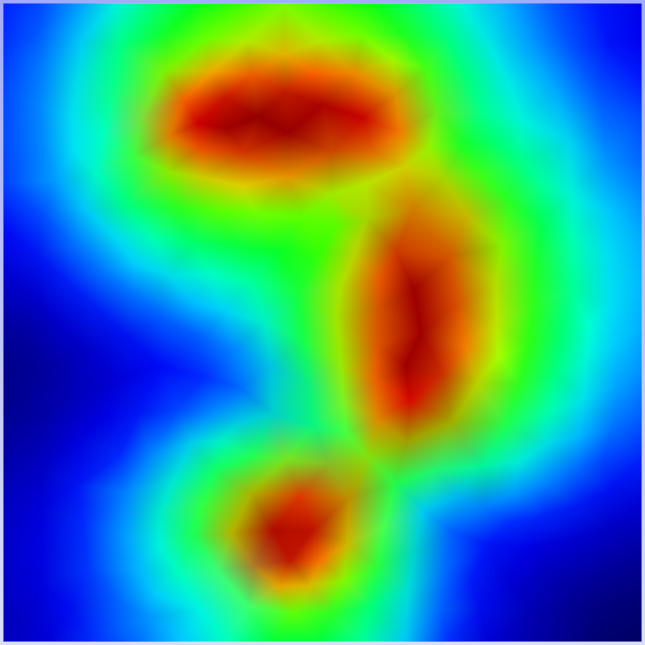}}
   \subfigure{\includegraphics[width=0.23\columnwidth]{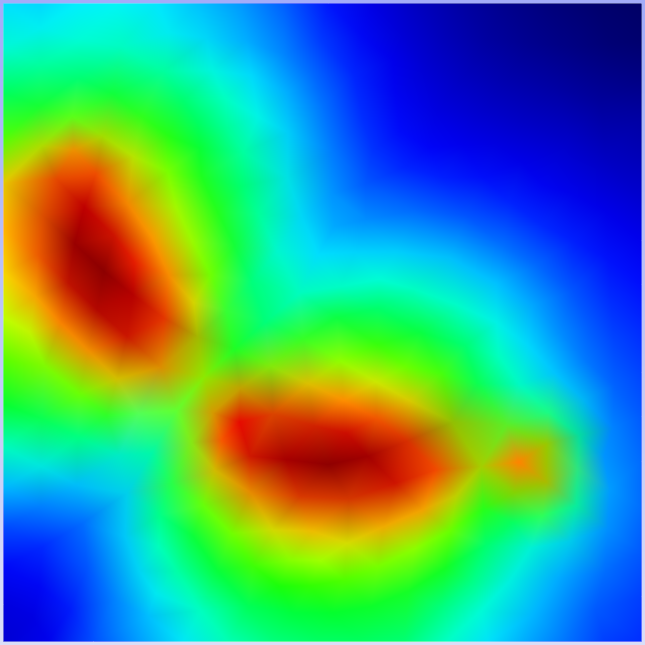}}
   \subfigure{\includegraphics[width=0.23\columnwidth]{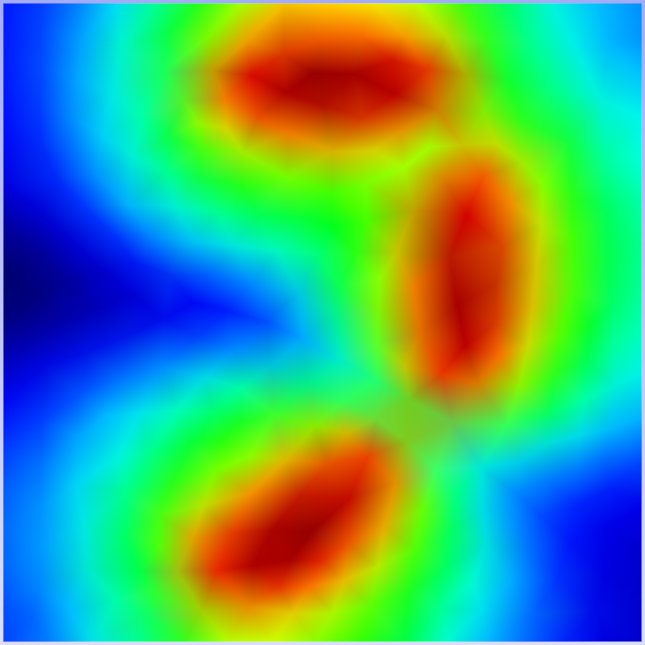}}
   \subfigure{\includegraphics[width=0.23\columnwidth]{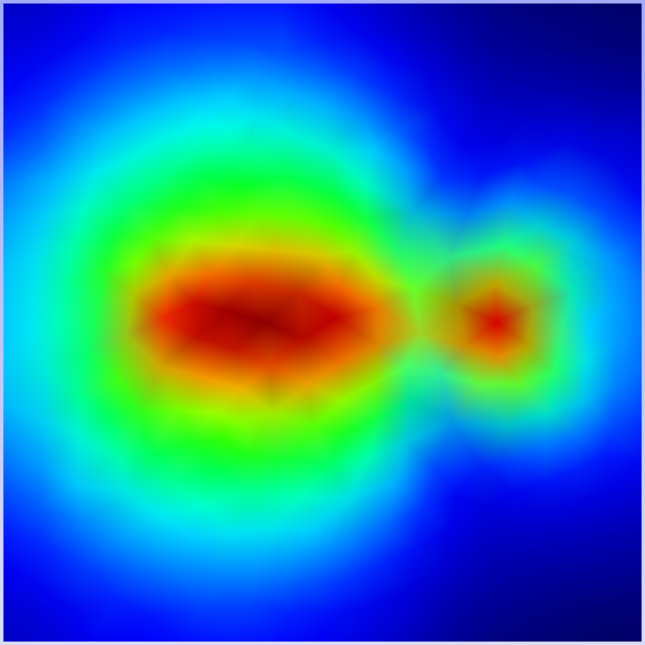}}
   \subfigure{\includegraphics[width=0.23\columnwidth]{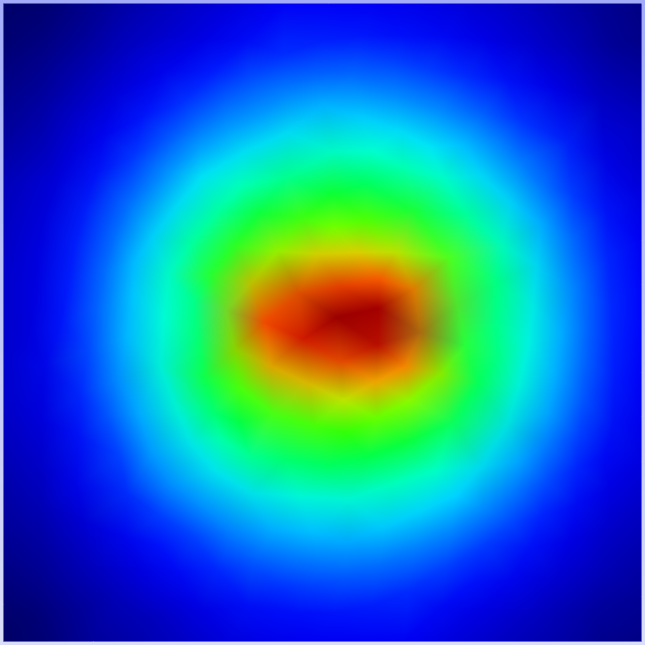}}
   \subfigure{\includegraphics[width=0.23\columnwidth]{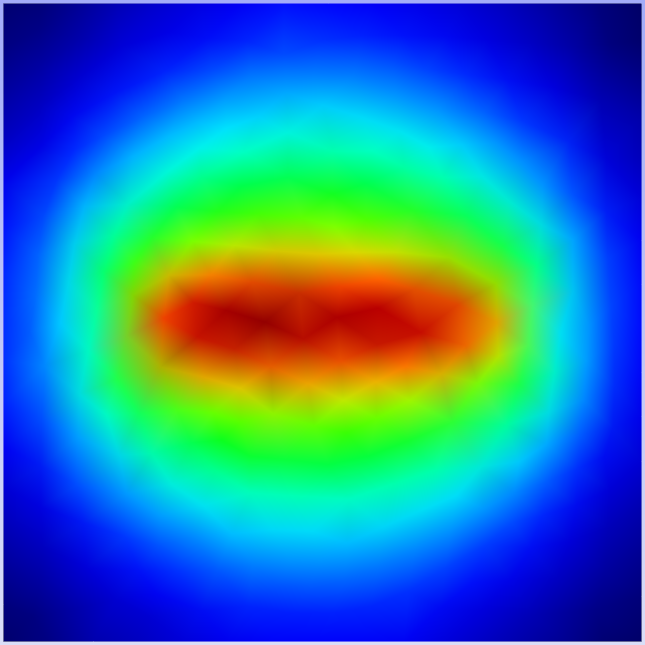}}
   \subfigure{\includegraphics[width=0.23\columnwidth]{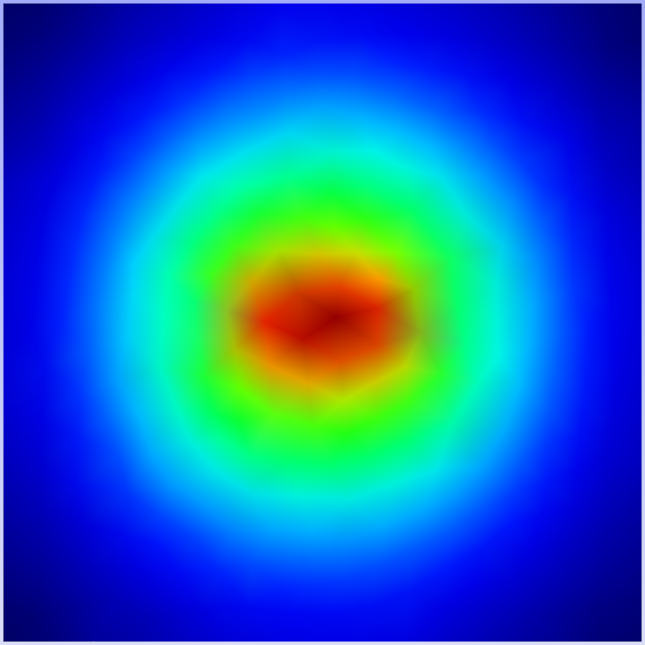}}
  \subfigure{\includegraphics[width=0.23\columnwidth]{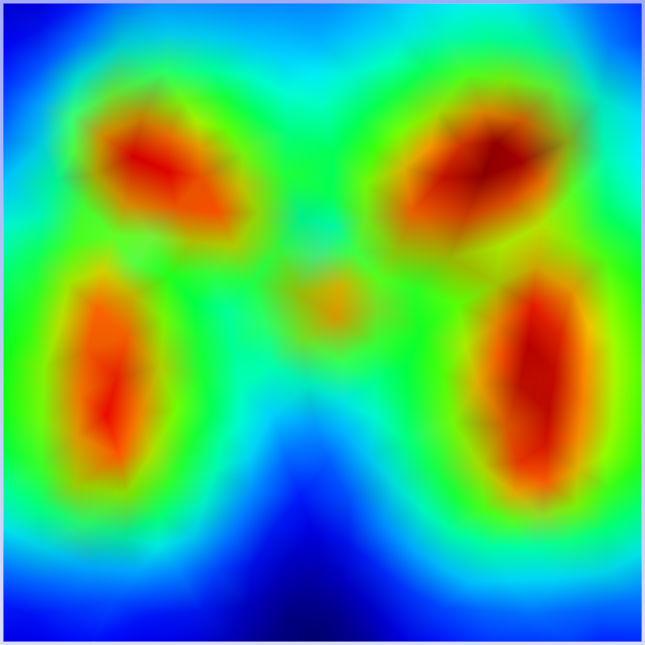}}
   \subfigure{\includegraphics[width=0.23\columnwidth]{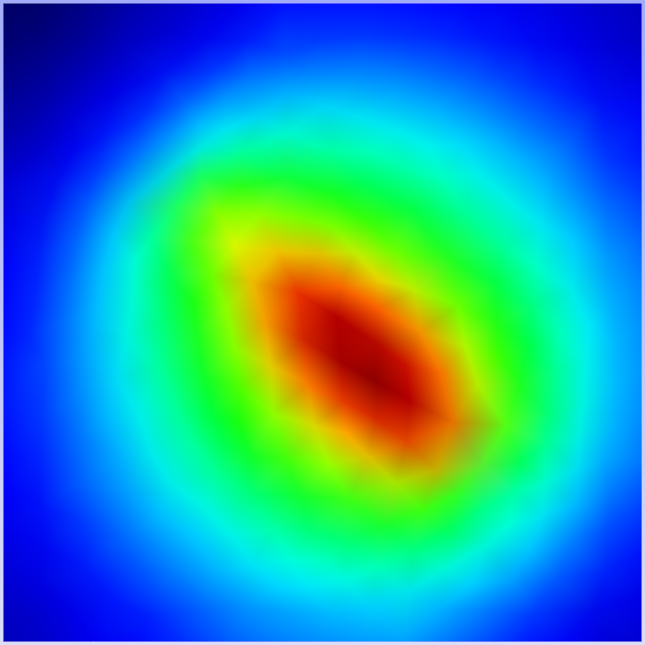}}
   \subfigure{\includegraphics[width=0.23\columnwidth]{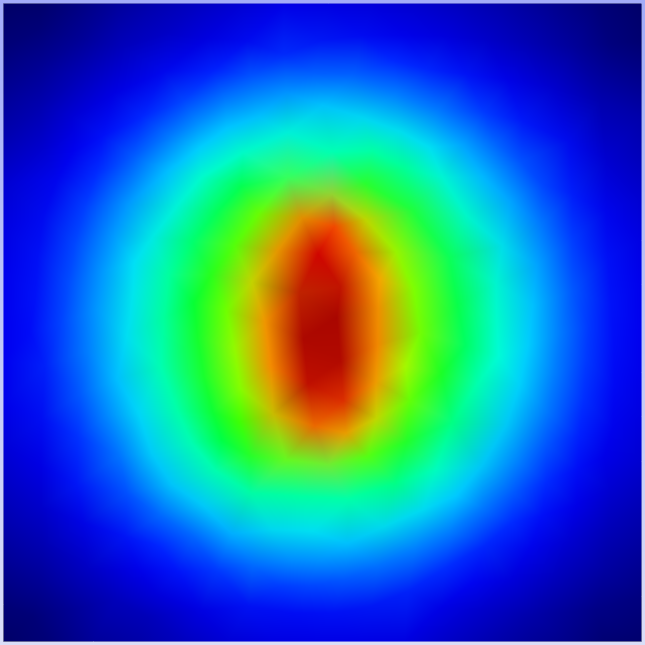}}
   \subfigure{\includegraphics[width=0.23\columnwidth]{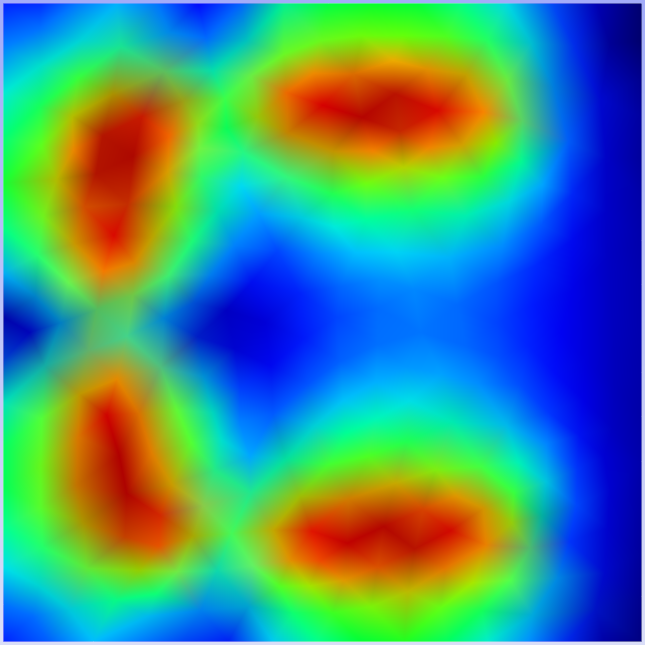}}	
   \subfigure{\includegraphics[width=0.23\columnwidth]{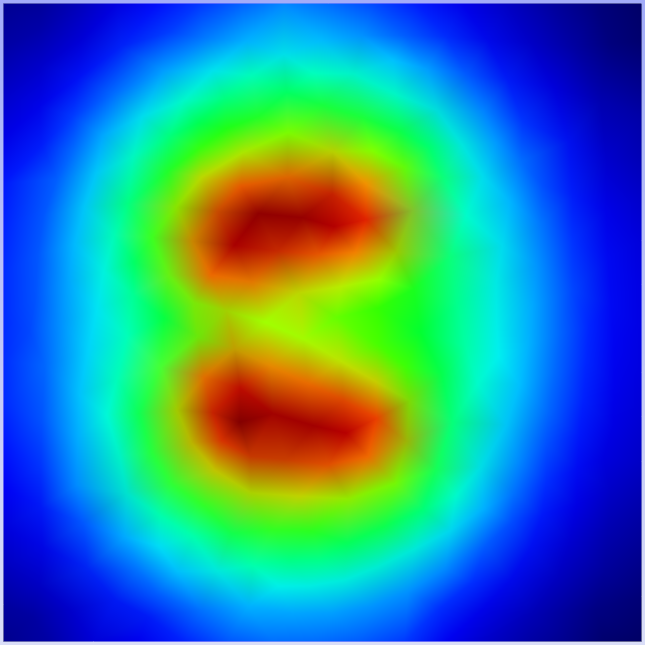}}
   \subfigure{\includegraphics[width=0.23\columnwidth]{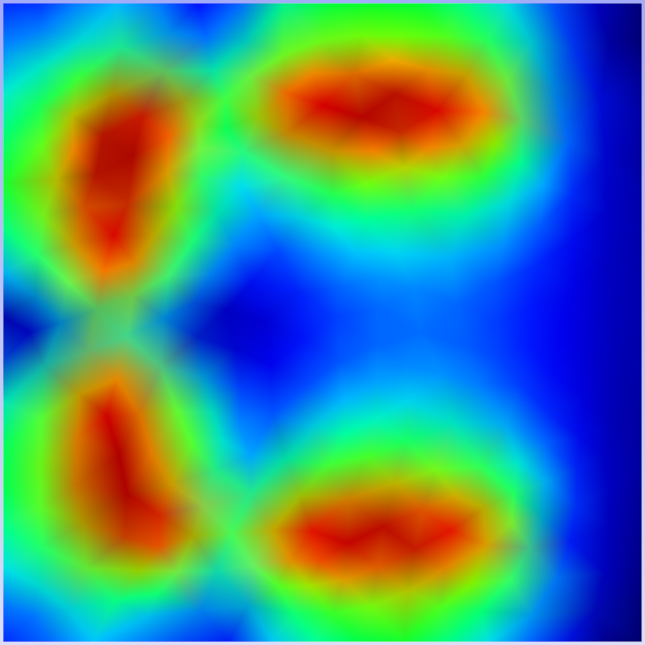}}
  \subfigure{\includegraphics[width=0.23\columnwidth]{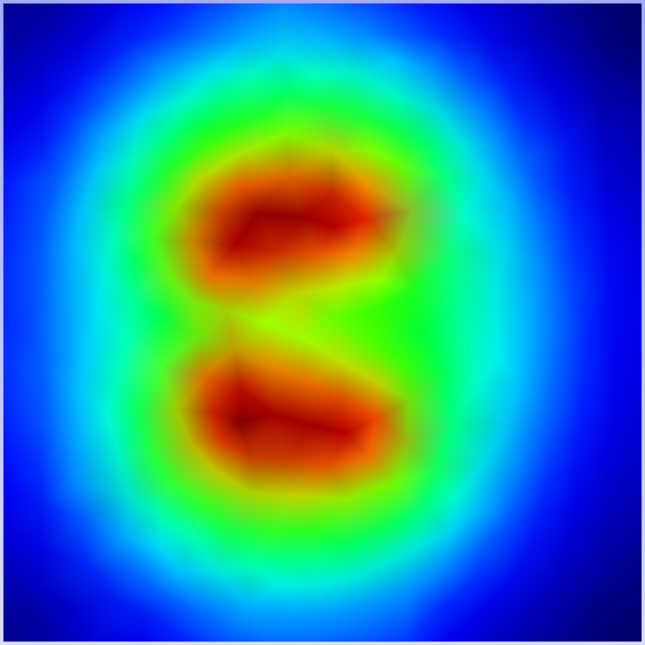}}
  \subfigure{\includegraphics[width=0.23\columnwidth]{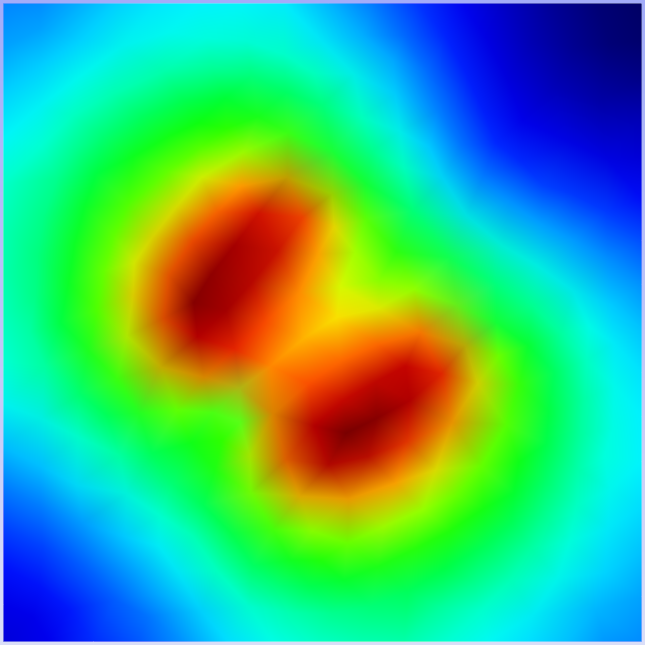}}
   \subfigure{\includegraphics[width=0.23\columnwidth]{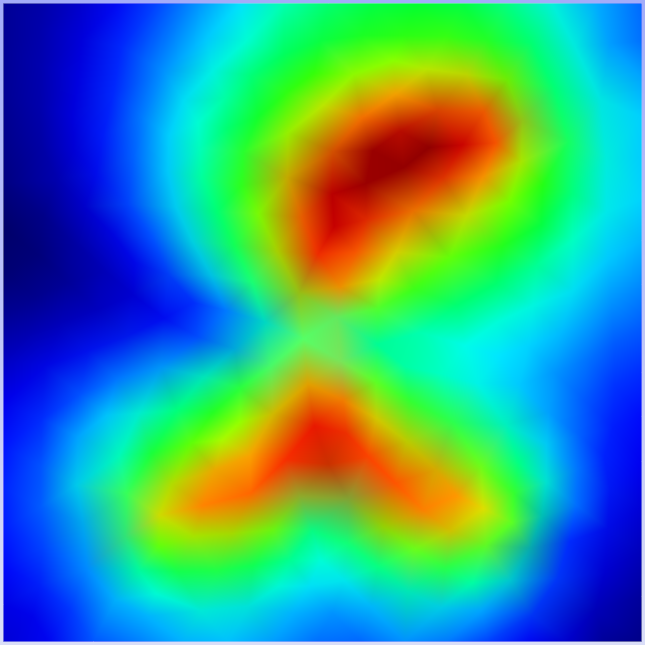}}
   \subfigure{\includegraphics[width=0.23\columnwidth]{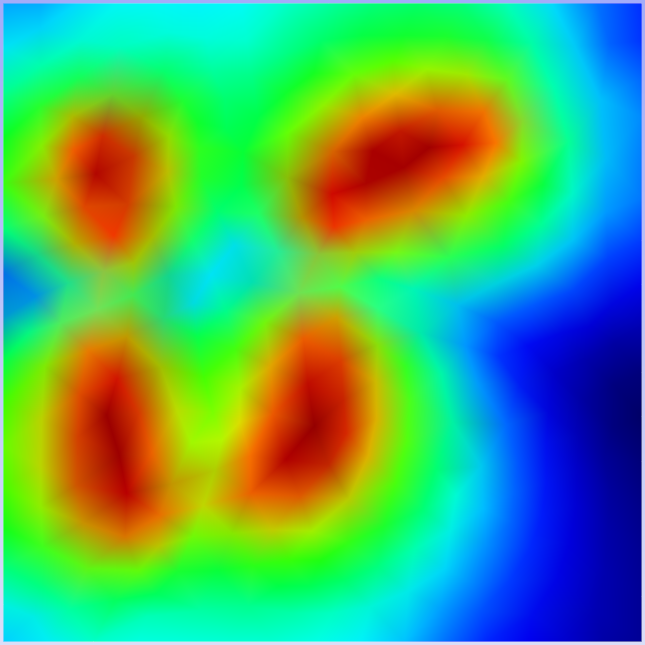}}
   \subfigure{\includegraphics[width=0.23\columnwidth]{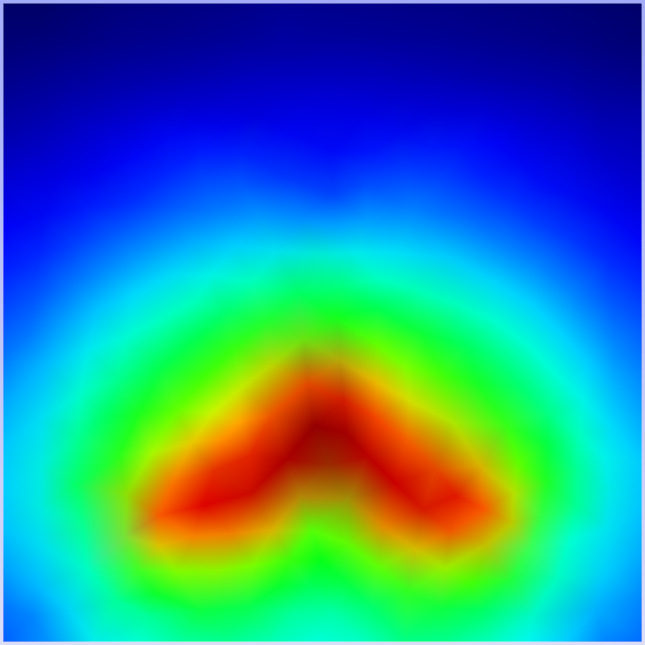}}
   \subfigure{\includegraphics[width=0.23\columnwidth]{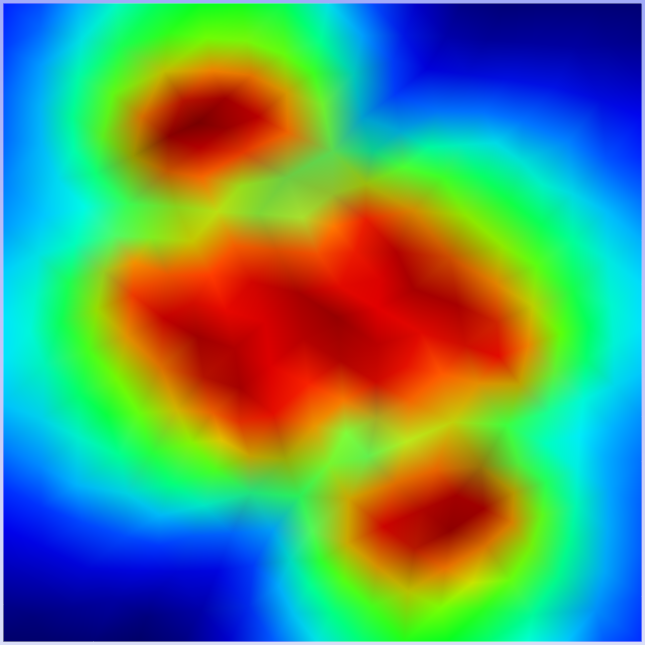}}
   \subfigure{\includegraphics[width=0.23\columnwidth]{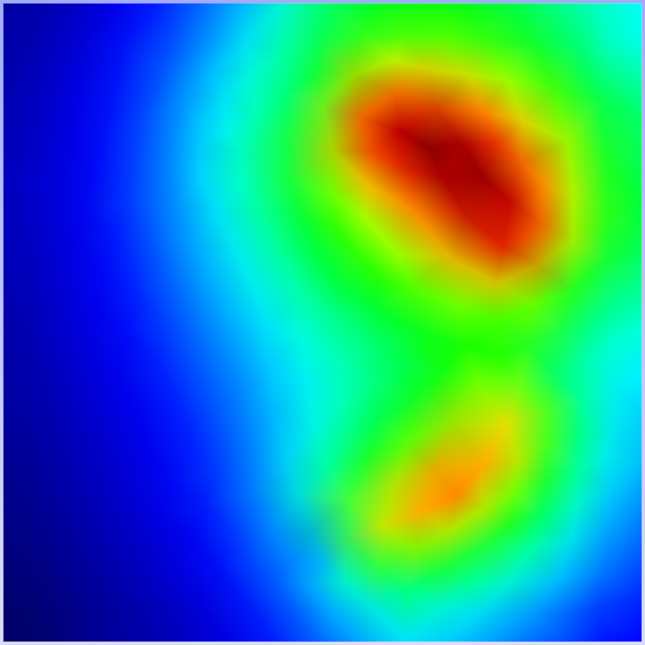}}
   \subfigure{\includegraphics[width=0.23\columnwidth]{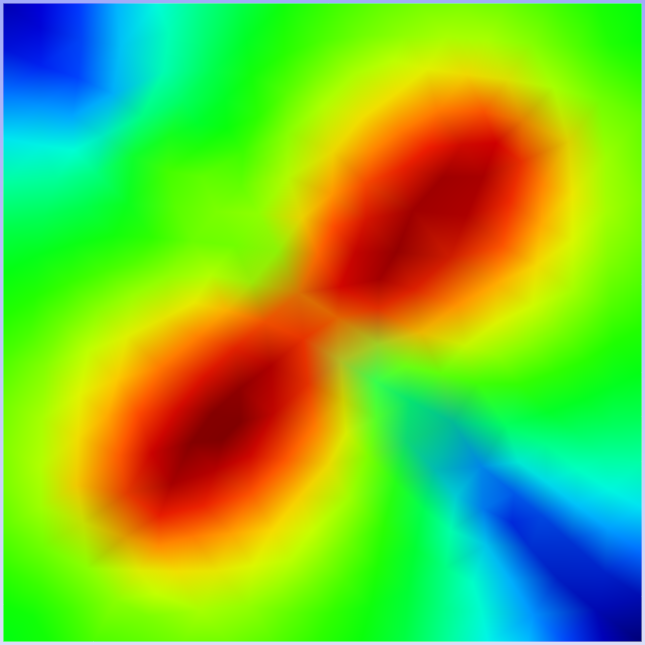}}
   \subfigure{\includegraphics[width=0.23\columnwidth]{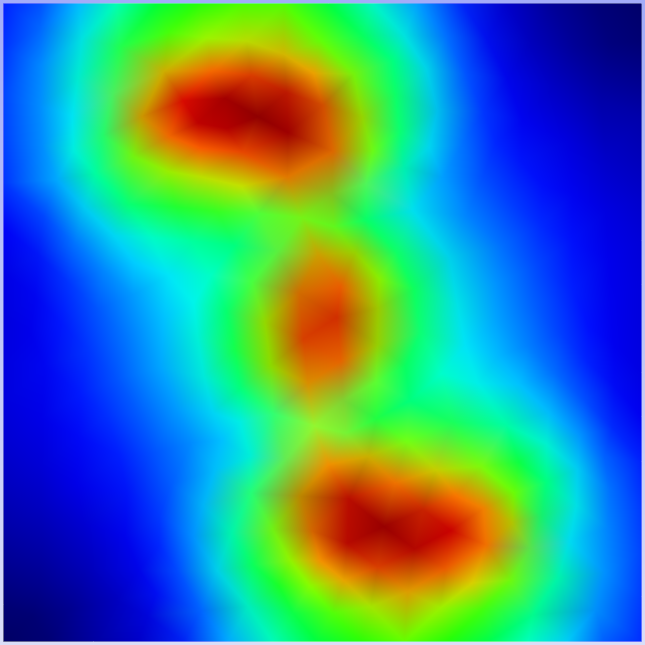}}
   \subfigure{\includegraphics[width=0.23\columnwidth]{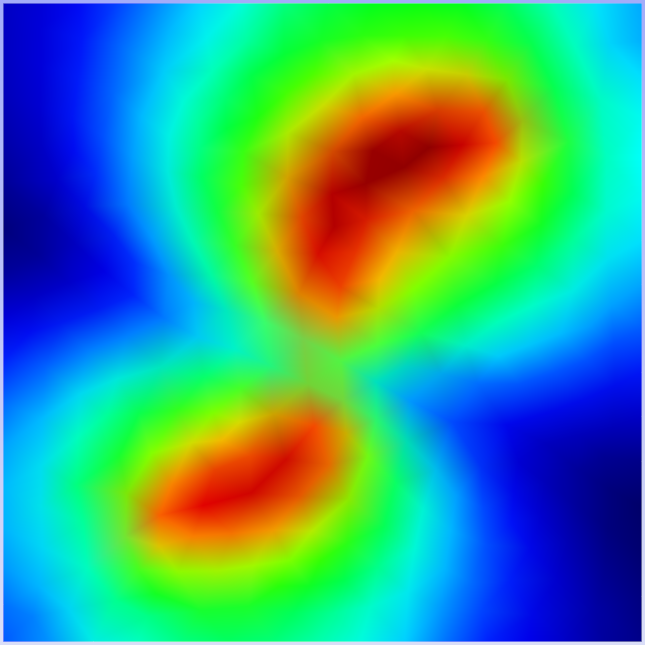}}
   \subfigure{\includegraphics[width=0.23\columnwidth]{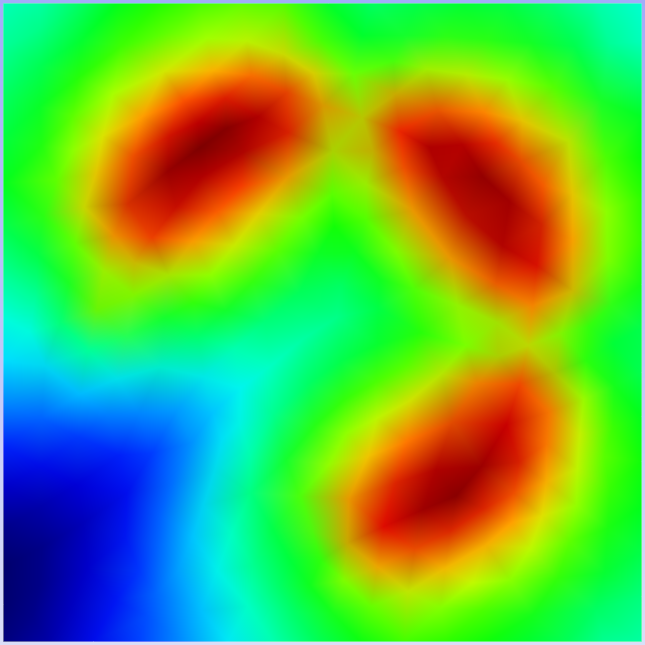}}
   \end{center} 
\caption{Magnitude of the radiated magnetic field from 32 radiating open wires with different shapes.} 
\label{Fig2} 
\end{figure}

\begin{figure}[!t] 
   \begin{center}
   \subfigure{\includegraphics[width=0.23\columnwidth]{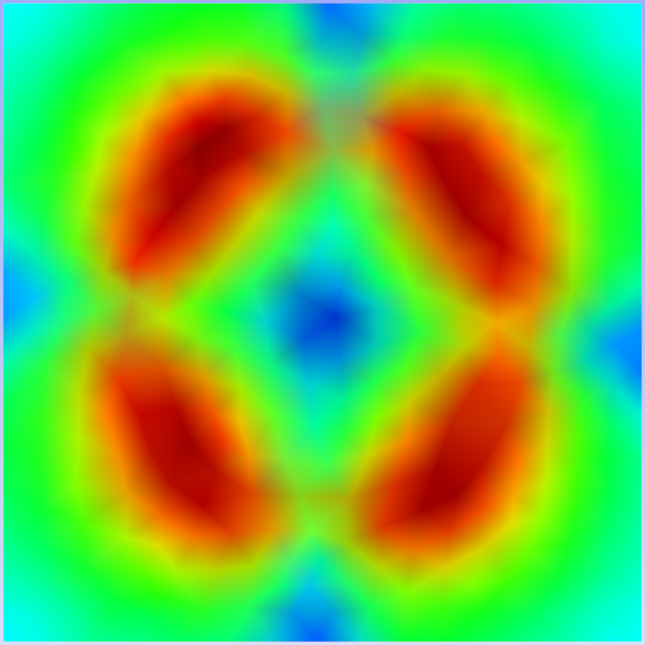}}
   \subfigure{\includegraphics[width=0.23\columnwidth]{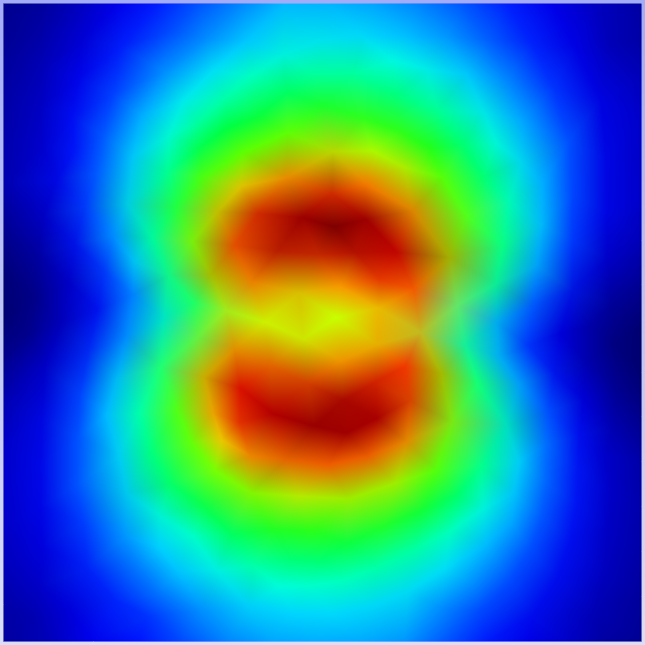}}
   \subfigure{\includegraphics[width=0.23\columnwidth]{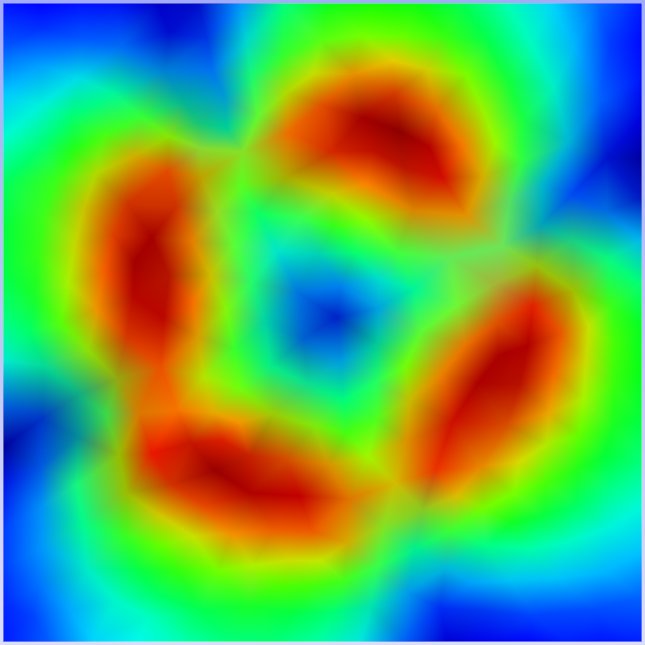}}
   \subfigure{\includegraphics[width=0.23\columnwidth]{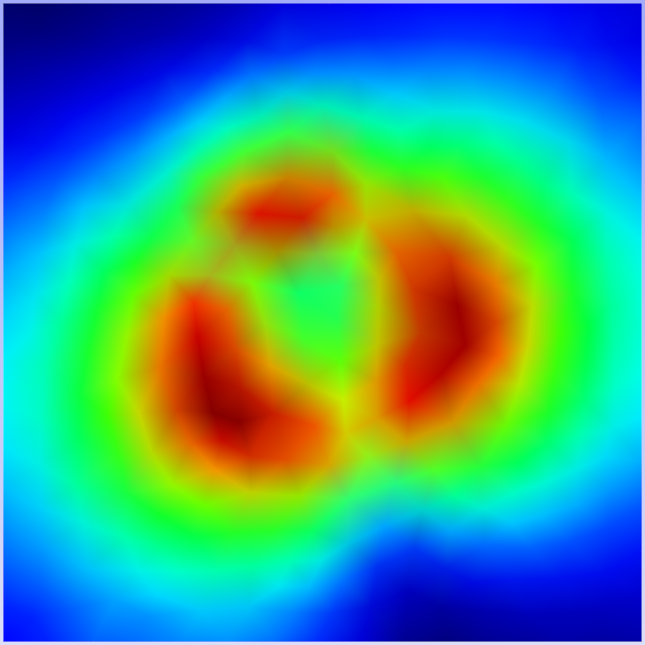}}
   \subfigure{\includegraphics[width=0.23\columnwidth]{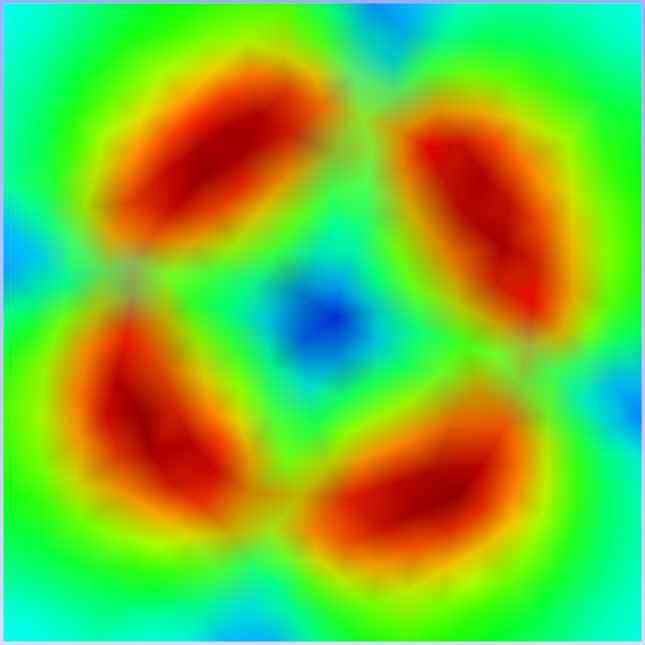}}
   \subfigure{\includegraphics[width=0.23\columnwidth]{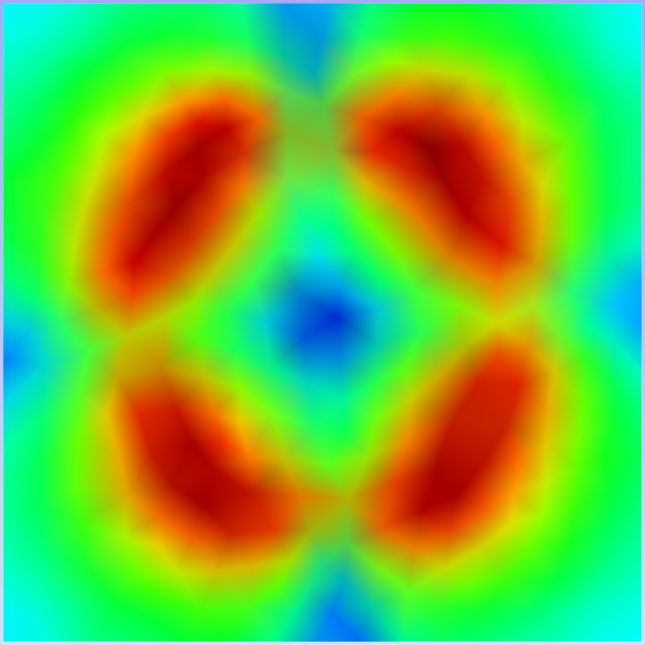}}
   \subfigure{\includegraphics[width=0.23\columnwidth]{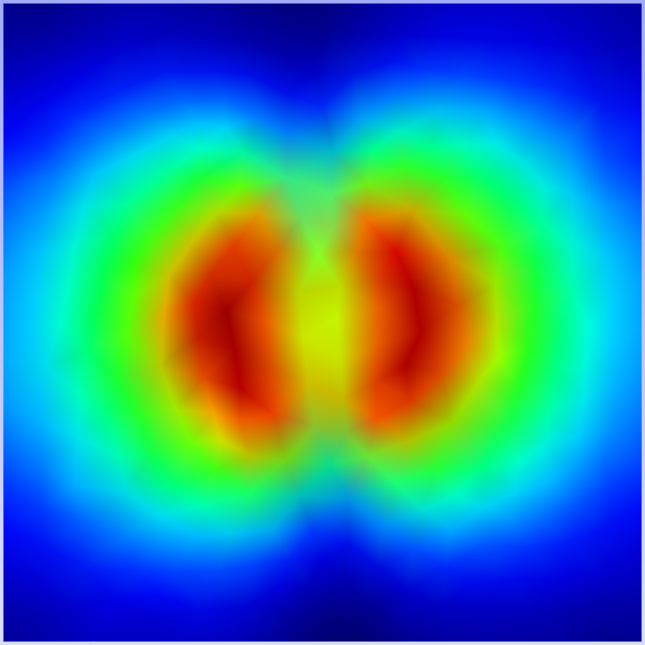}}
   \subfigure{\includegraphics[width=0.23\columnwidth]{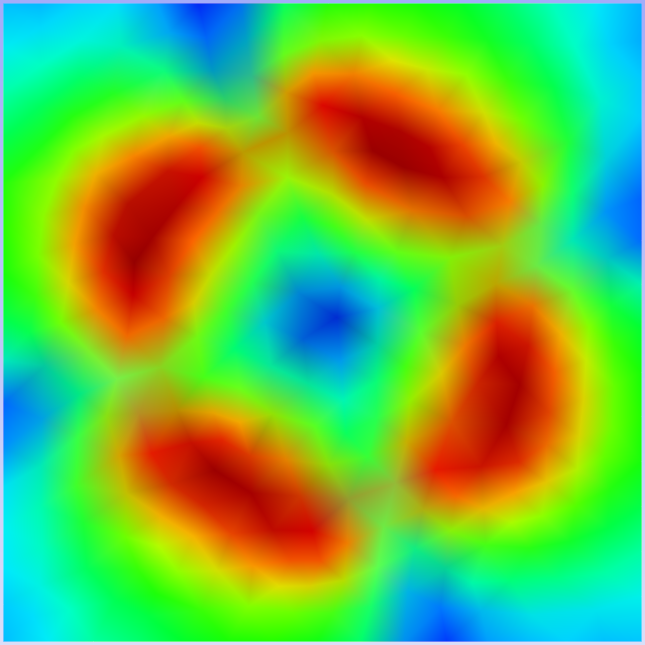}}
   \subfigure{\includegraphics[width=0.23\columnwidth]{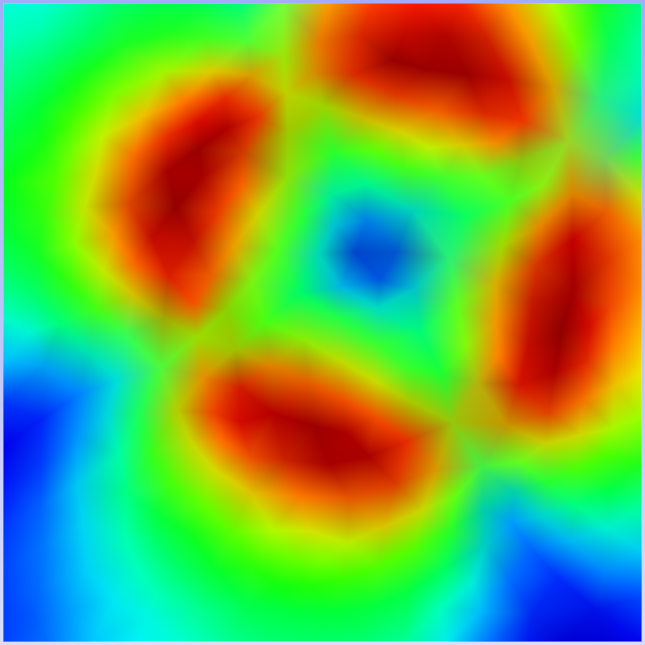}}
   \subfigure{\includegraphics[width=0.23\columnwidth]{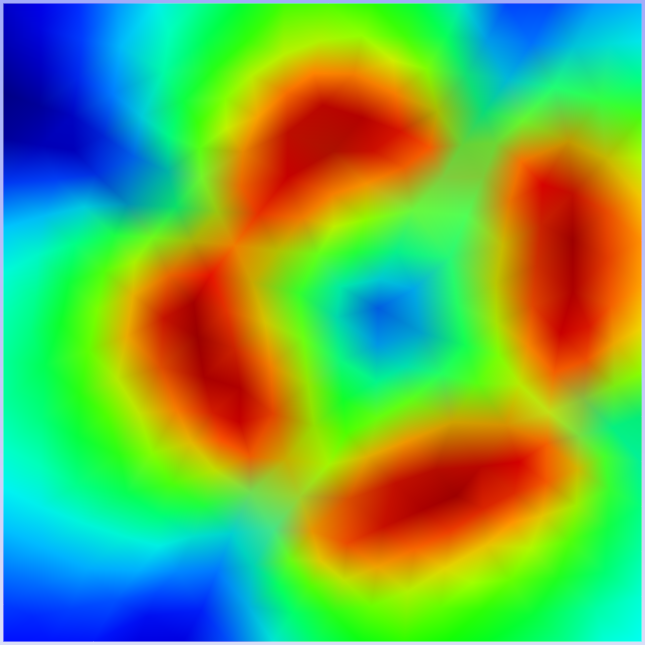}}
   \subfigure{\includegraphics[width=0.23\columnwidth]{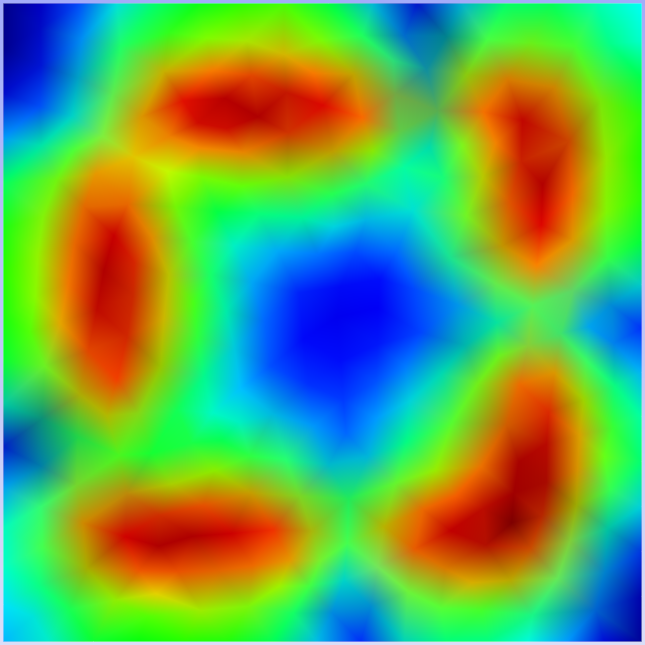}}
   \subfigure{\includegraphics[width=0.23\columnwidth]{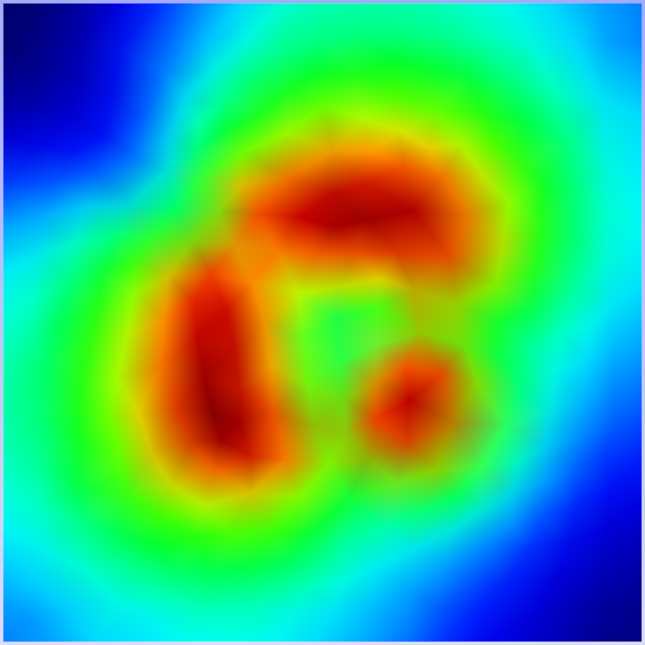}}
   \subfigure{\includegraphics[width=0.23\columnwidth]{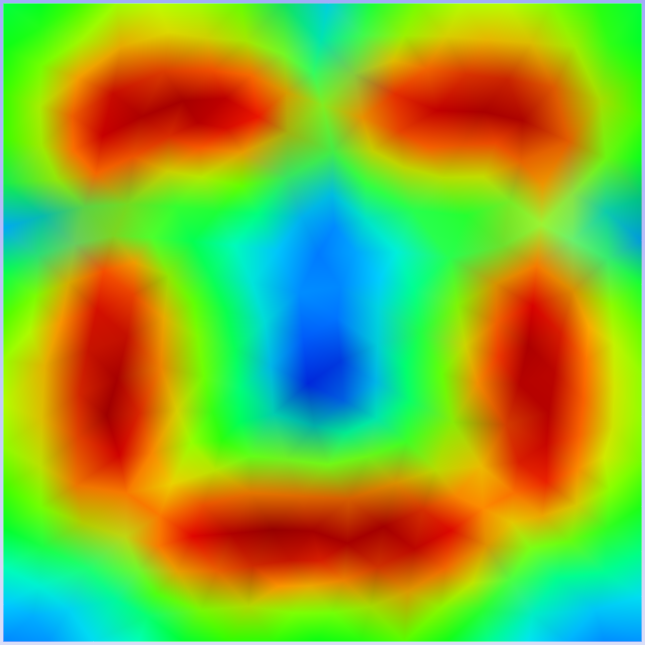}}
   \subfigure{\includegraphics[width=0.23\columnwidth]{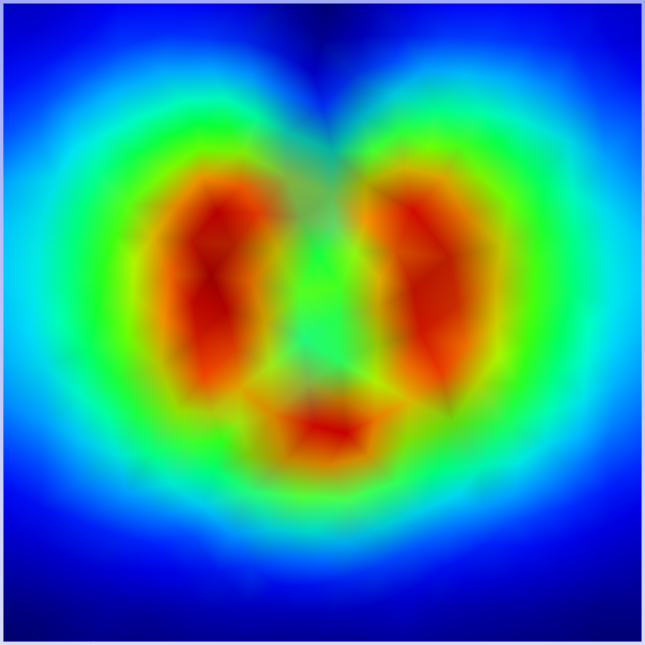}}
   \subfigure{\includegraphics[width=0.23\columnwidth]{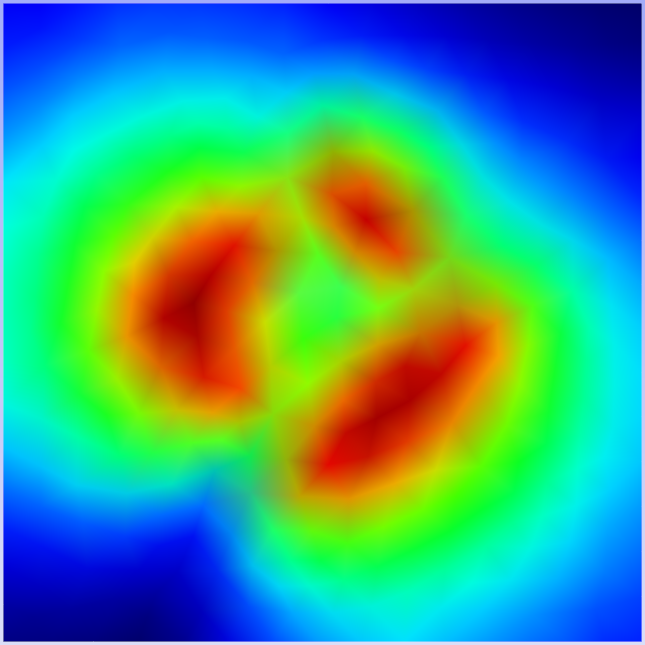}}
   \subfigure{\includegraphics[width=0.23\columnwidth]{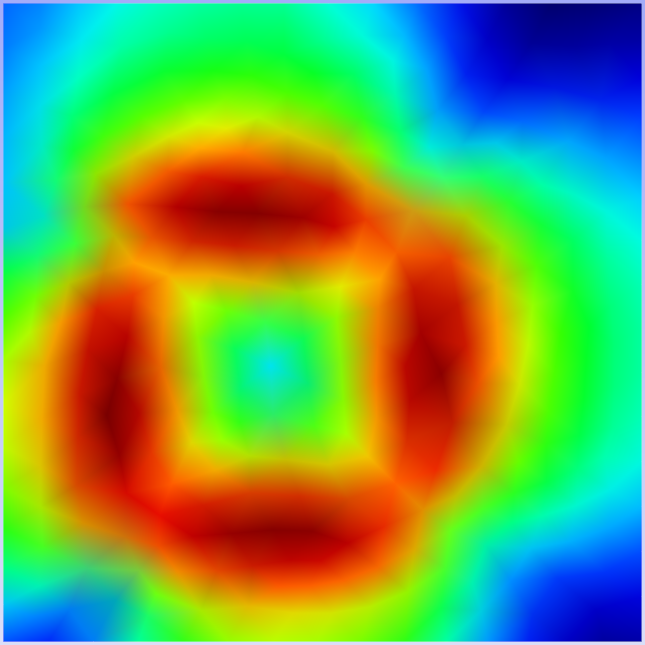}}
   \subfigure{\includegraphics[width=0.23\columnwidth]{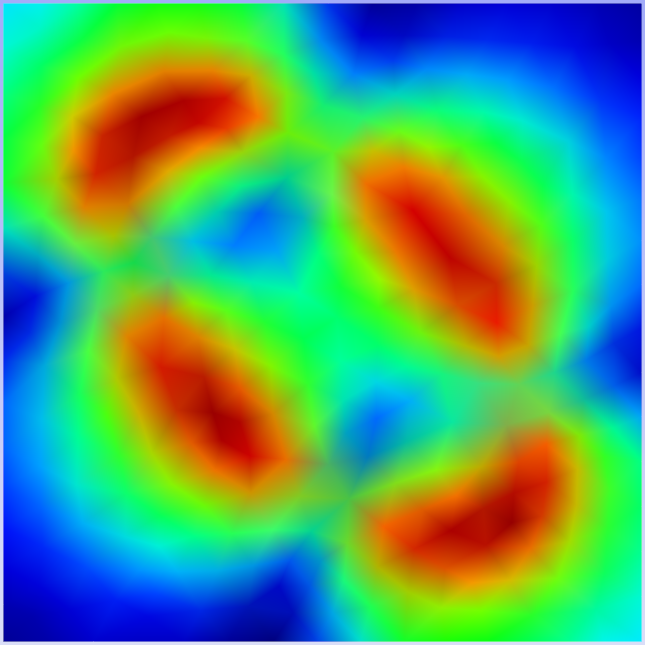}}
   \subfigure{\includegraphics[width=0.23\columnwidth]{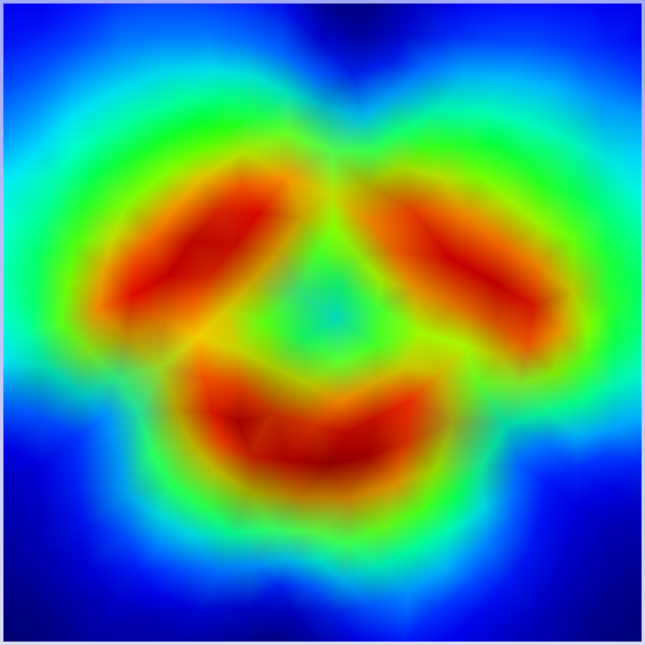}}
   \subfigure{\includegraphics[width=0.23\columnwidth]{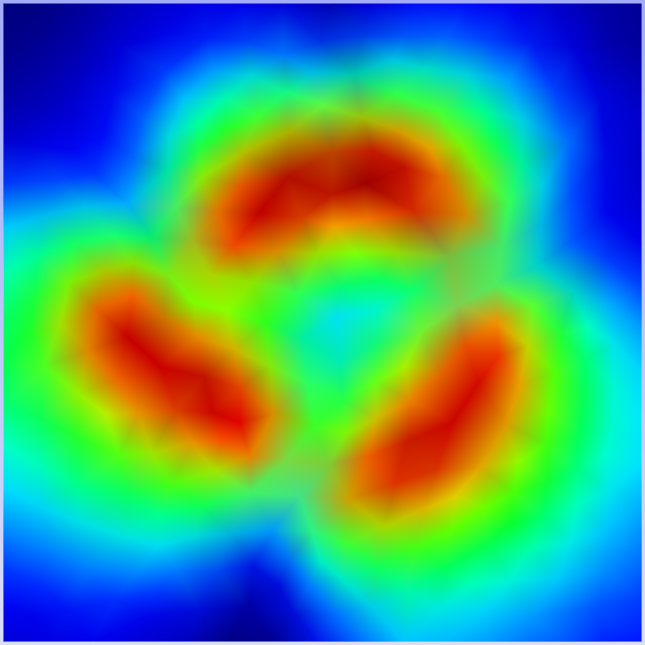}}	
   \subfigure{\includegraphics[width=0.23\columnwidth]{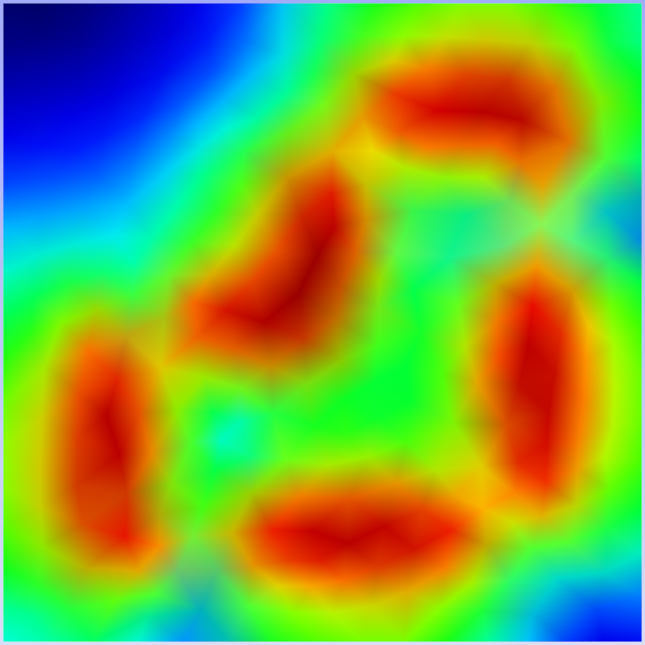}}
   \subfigure{\includegraphics[width=0.23\columnwidth]{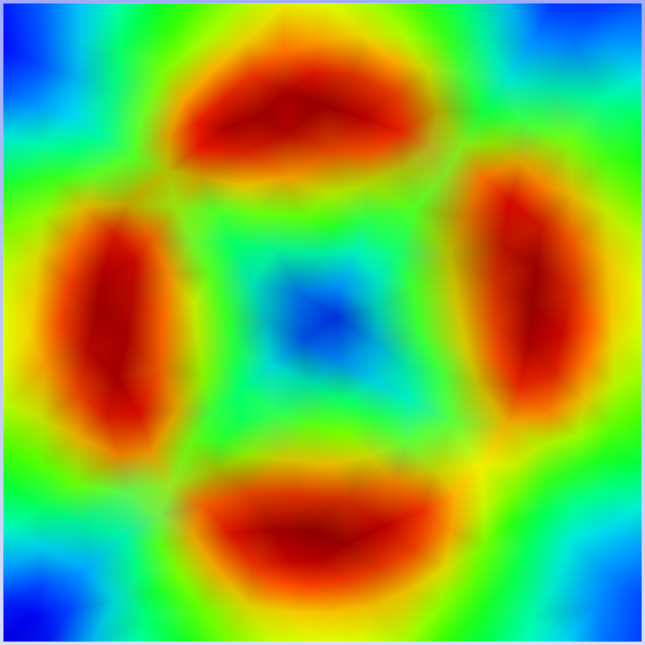}}
   \subfigure{\includegraphics[width=0.23\columnwidth]{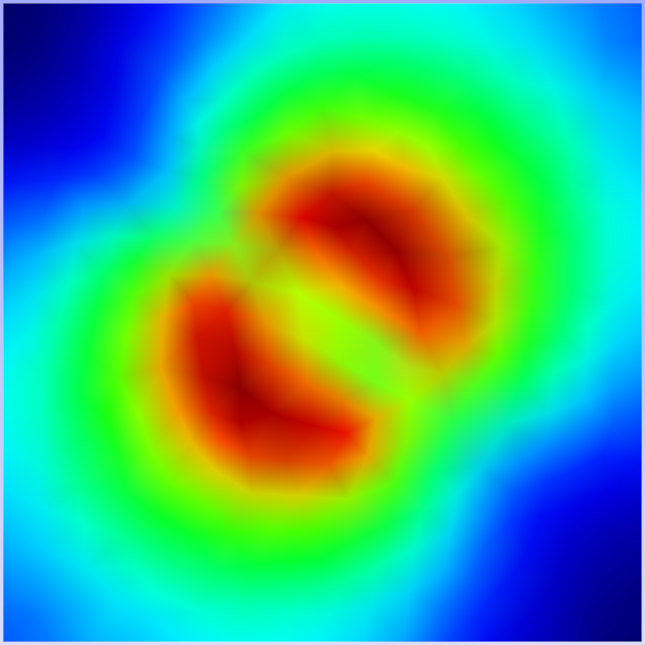}}
   \subfigure{\includegraphics[width=0.23\columnwidth]{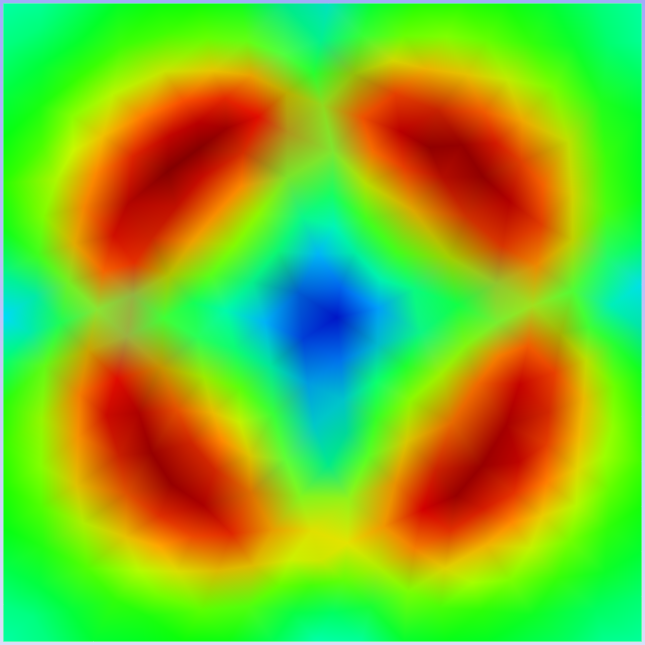}}
   \subfigure{\includegraphics[width=0.23\columnwidth]{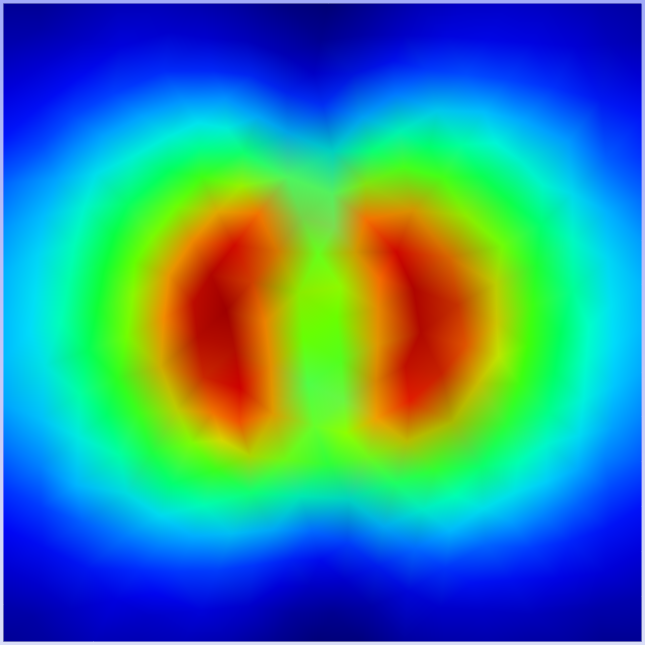}}
   \subfigure{\includegraphics[width=0.23\columnwidth]{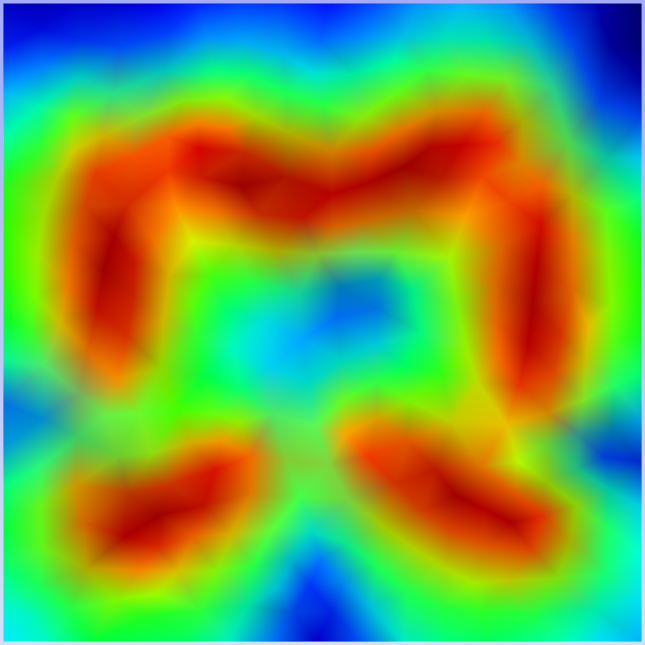}}
   \subfigure{\includegraphics[width=0.23\columnwidth]{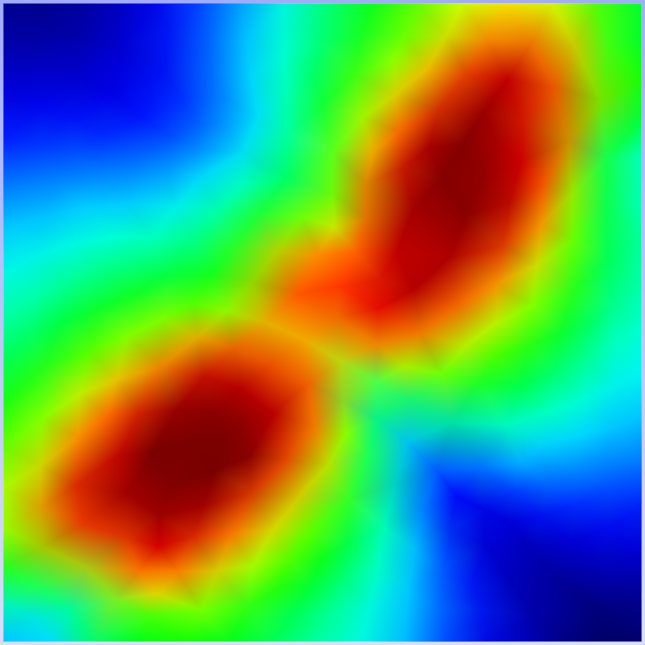}}
   \subfigure{\includegraphics[width=0.23\columnwidth]{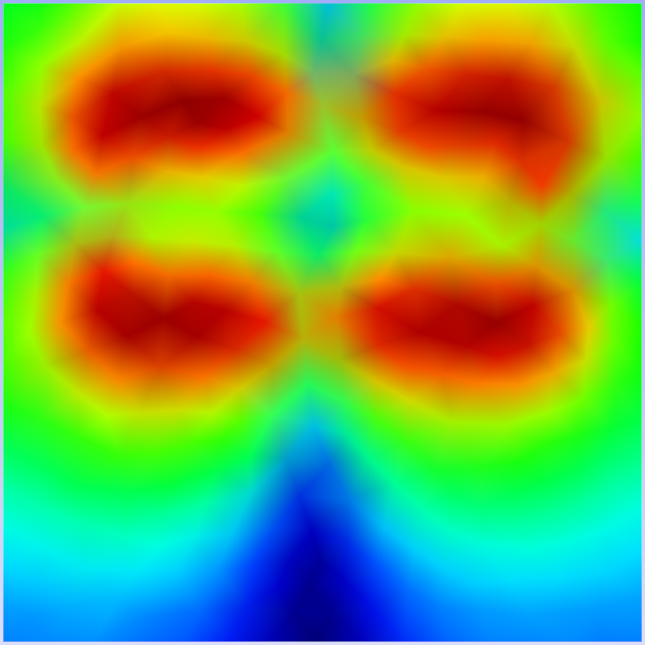}}
   \subfigure{\includegraphics[width=0.23\columnwidth]{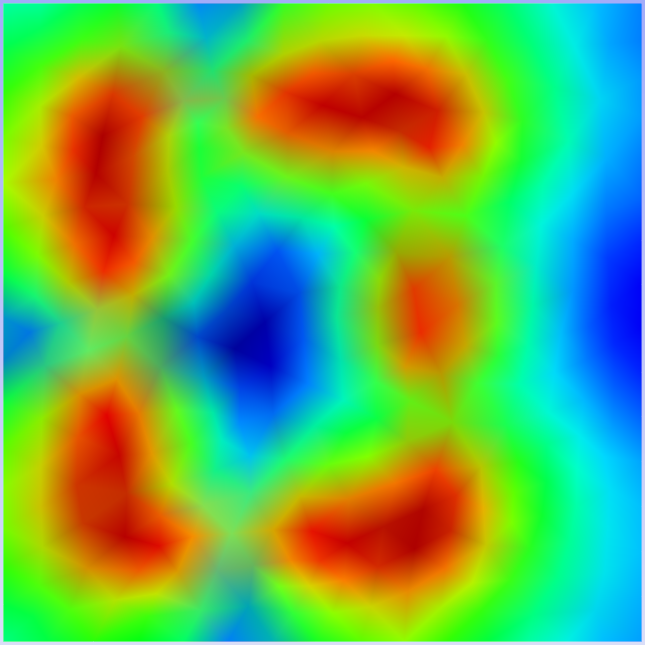}}
   \subfigure{\includegraphics[width=0.23\columnwidth]{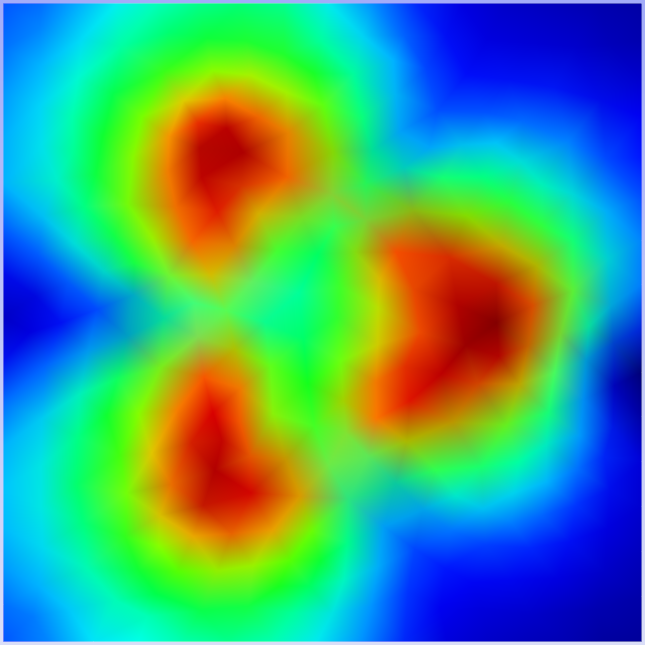}}
   \subfigure{\includegraphics[width=0.23\columnwidth]{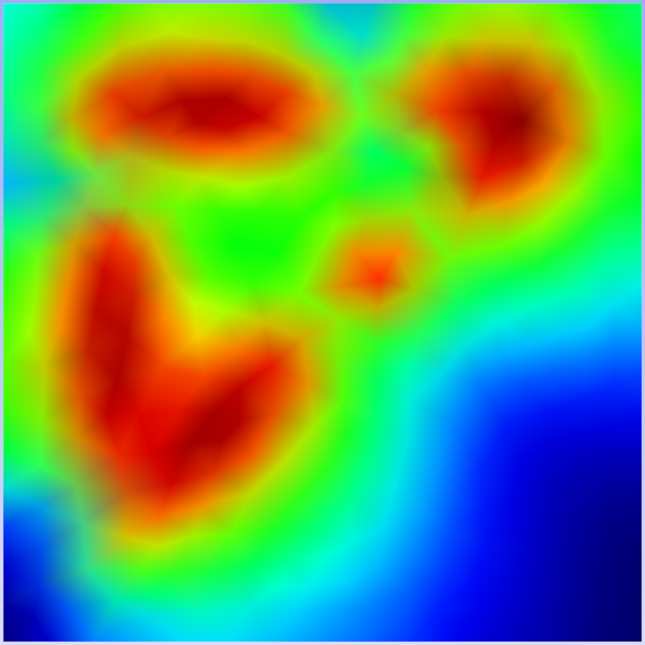}}
   \subfigure{\includegraphics[width=0.23\columnwidth]{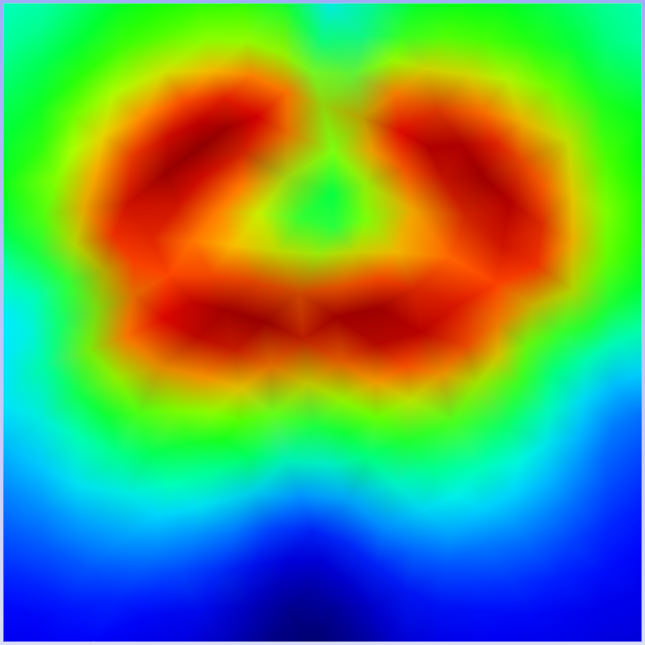}}
   \subfigure{\includegraphics[width=0.23\columnwidth]{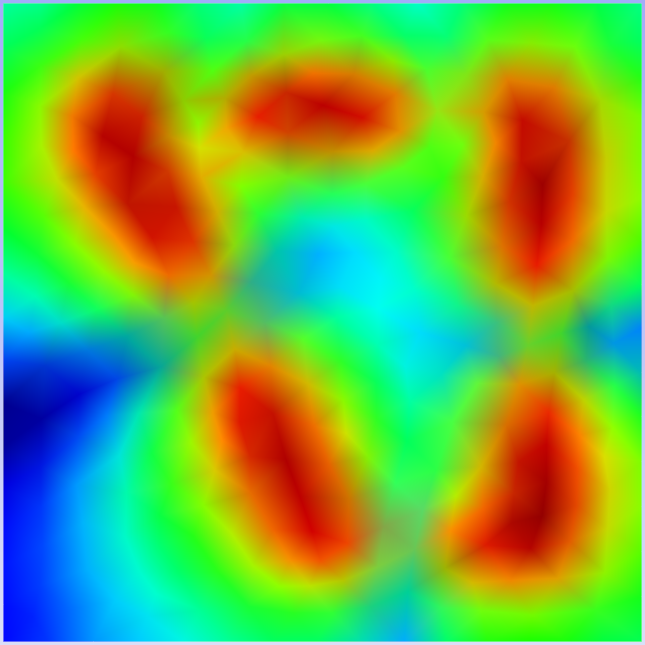}}
   \end{center} 
\caption{Magnitude of the radiated electric field from 32 radiating closed wires with different shapes.} 
\label{Fig3} 
\end{figure}

\begin{figure}[!t] 
   \begin{center}
   \subfigure{\includegraphics[width=0.23\columnwidth]{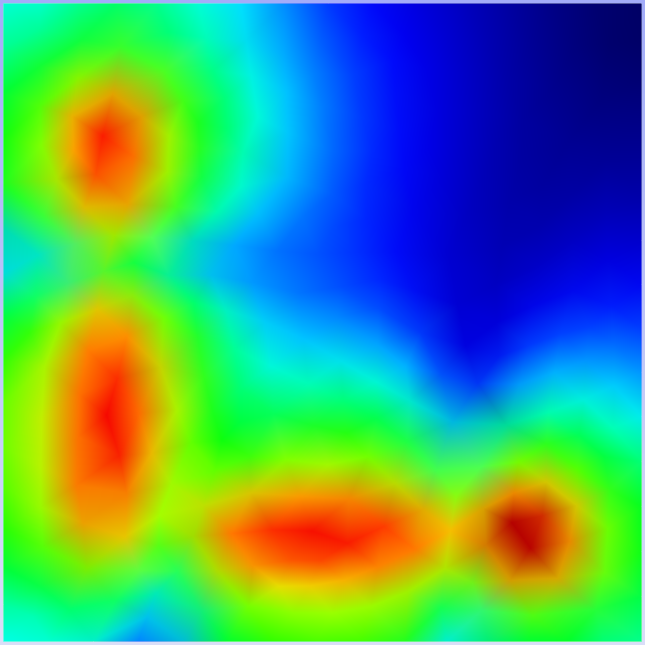}}
   \subfigure{\includegraphics[width=0.23\columnwidth]{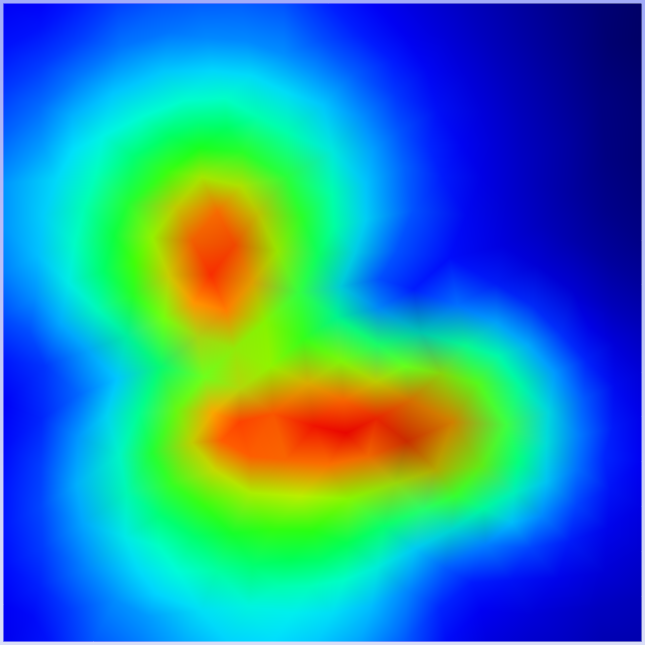}}
   \subfigure{\includegraphics[width=0.23\columnwidth]{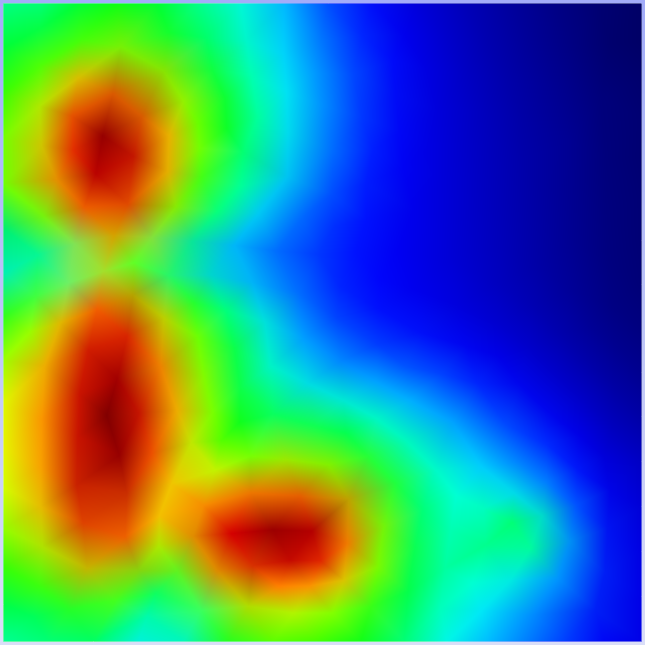}}
   \subfigure{\includegraphics[width=0.23\columnwidth]{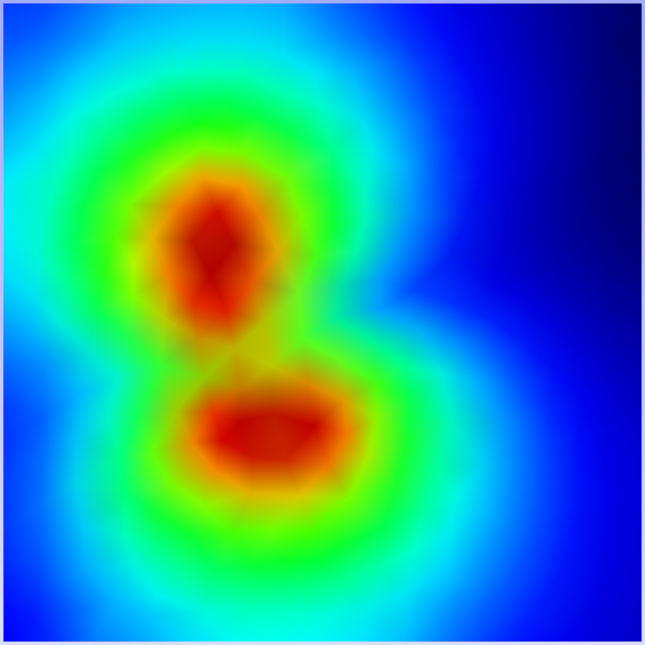}}
   \subfigure{\includegraphics[width=0.23\columnwidth]{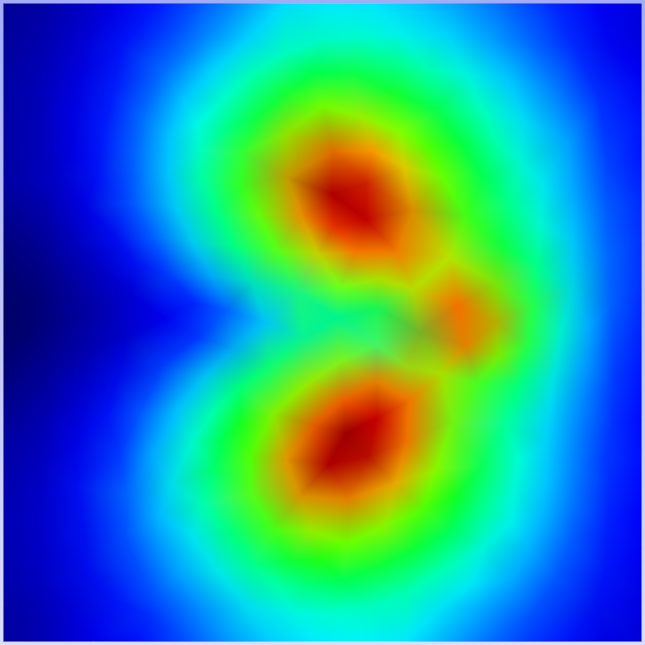}}
   \subfigure{\includegraphics[width=0.23\columnwidth]{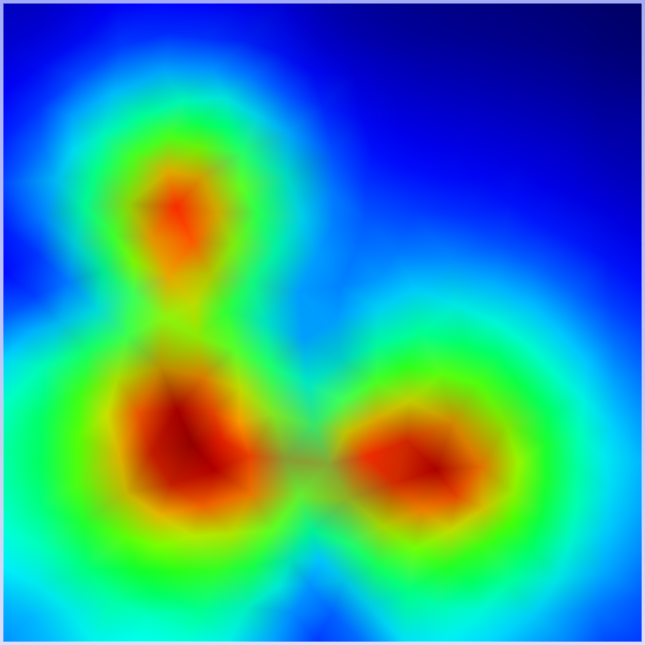}}
   \subfigure{\includegraphics[width=0.23\columnwidth]{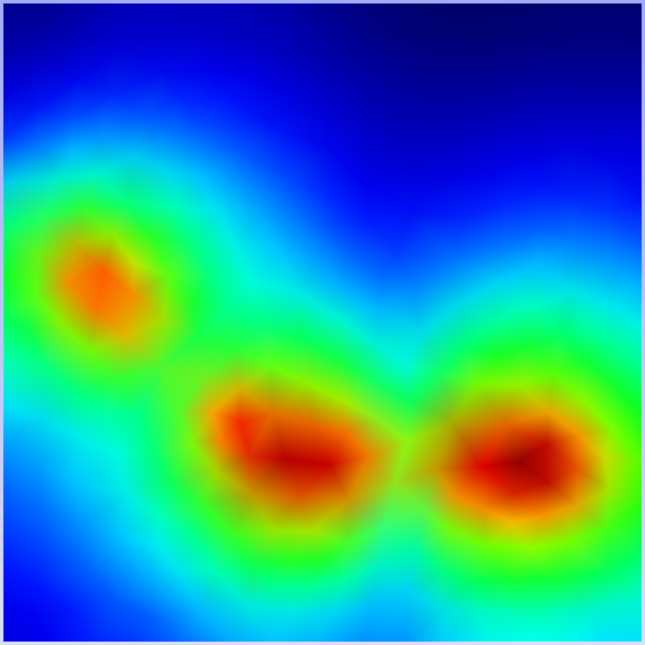}}
   \subfigure{\includegraphics[width=0.23\columnwidth]{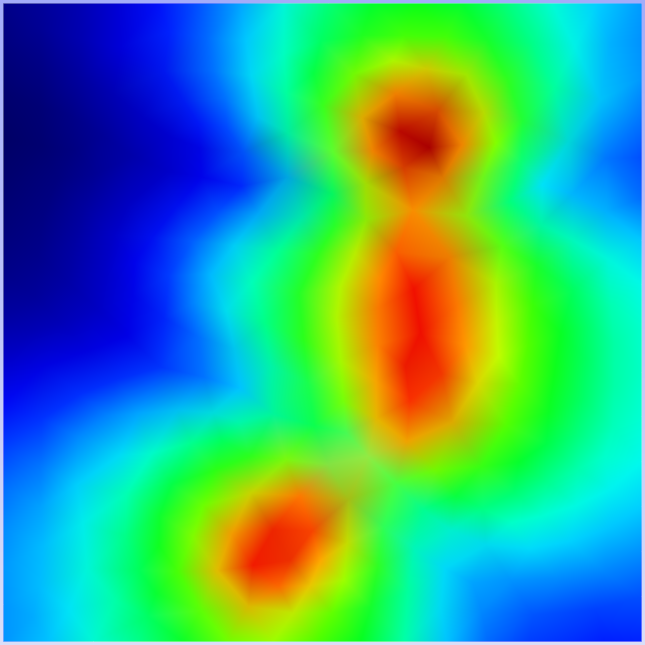}}
   \subfigure{\includegraphics[width=0.23\columnwidth]{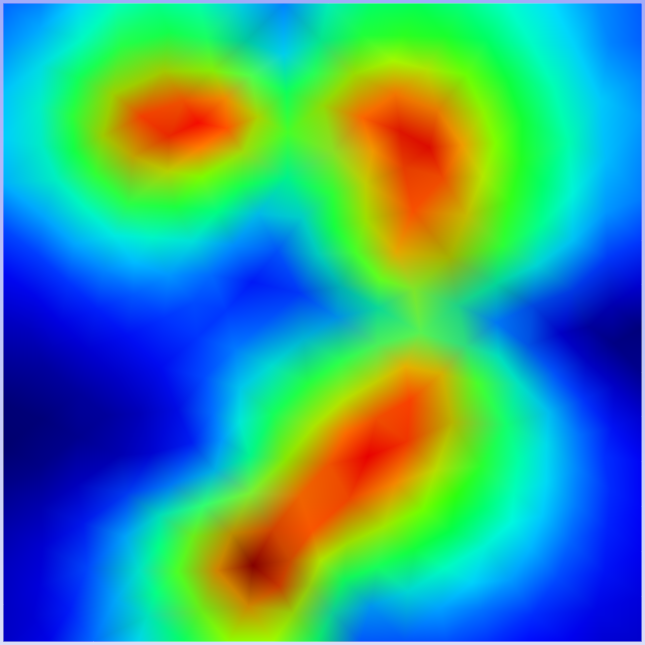}}
   \subfigure{\includegraphics[width=0.23\columnwidth]{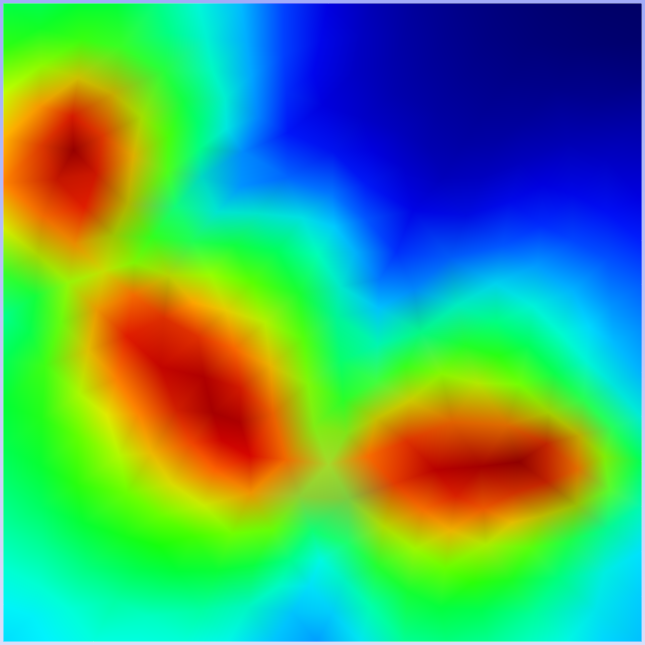}}
   \subfigure{\includegraphics[width=0.23\columnwidth]{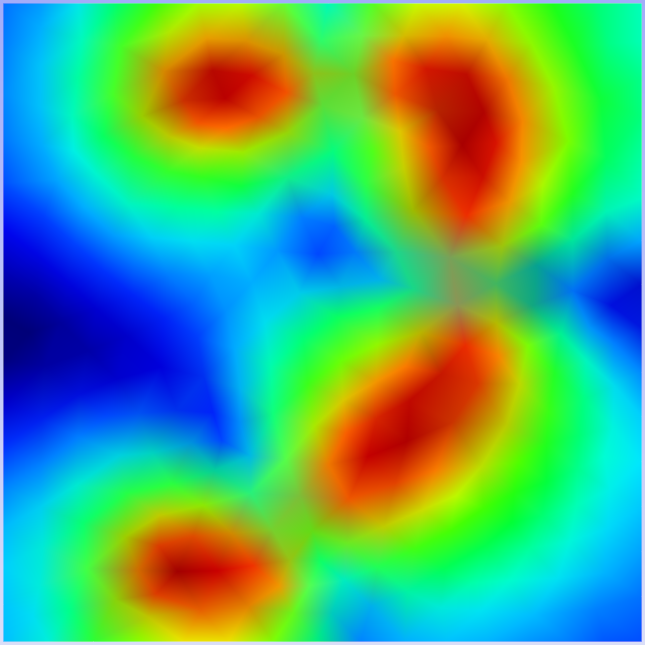}}
   \subfigure{\includegraphics[width=0.23\columnwidth]{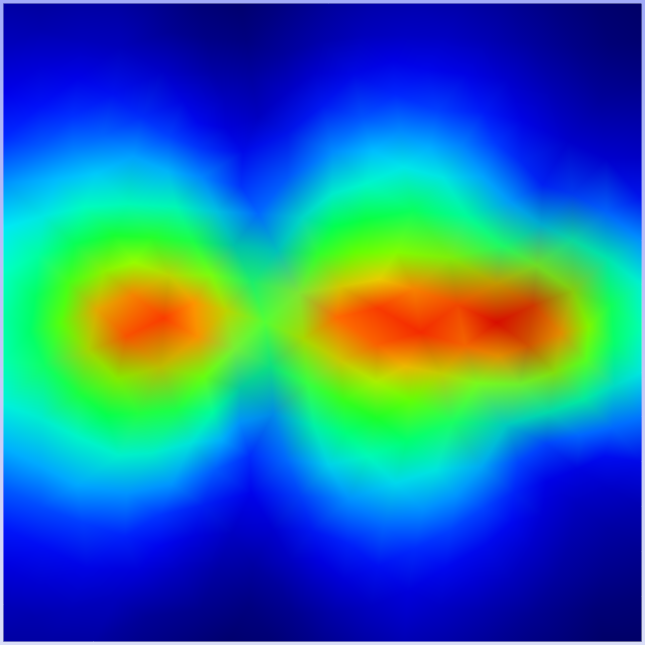}}
   \subfigure{\includegraphics[width=0.23\columnwidth]{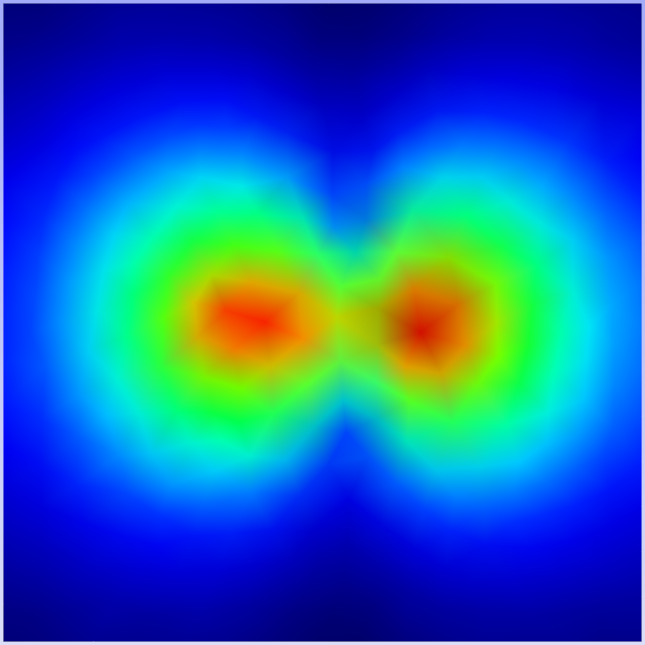}}
   \subfigure{\includegraphics[width=0.23\columnwidth]{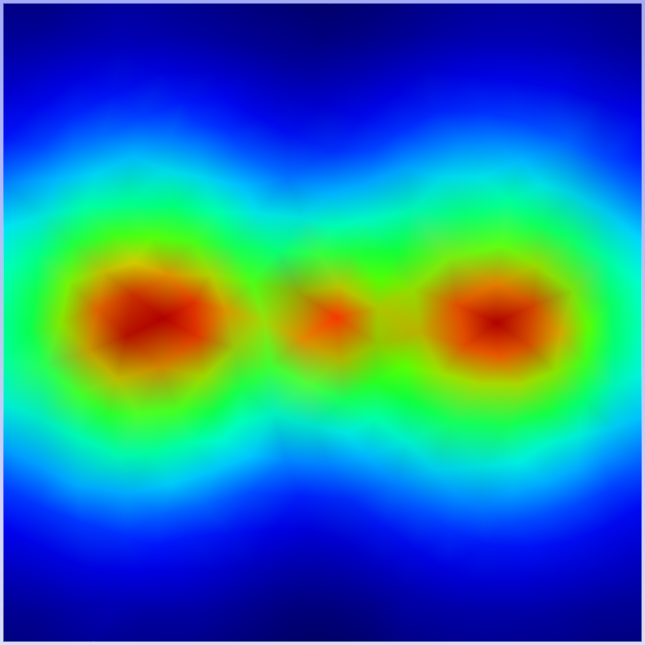}}
   \subfigure{\includegraphics[width=0.23\columnwidth]{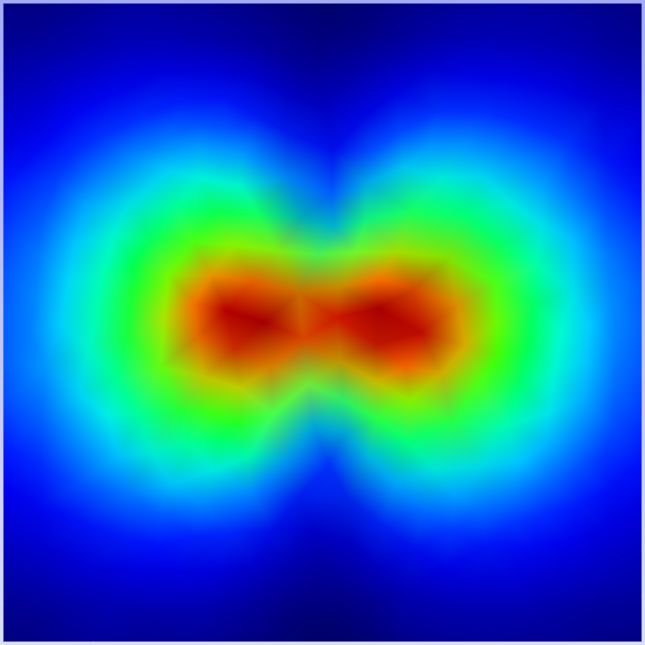}}
   \subfigure{\includegraphics[width=0.23\columnwidth]{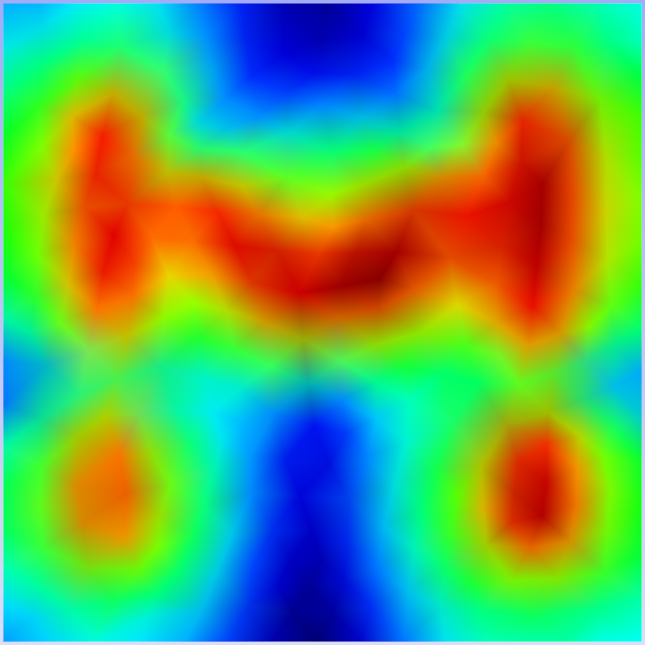}}
   \subfigure{\includegraphics[width=0.23\columnwidth]{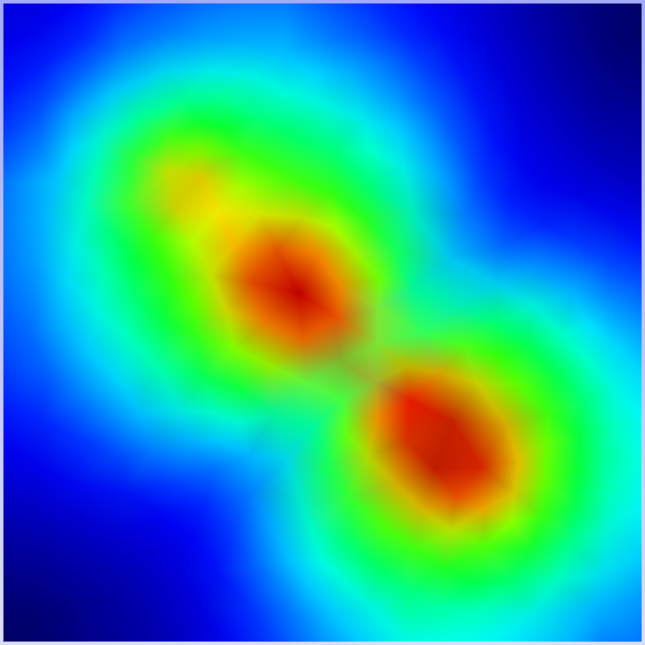}}
   \subfigure{\includegraphics[width=0.23\columnwidth]{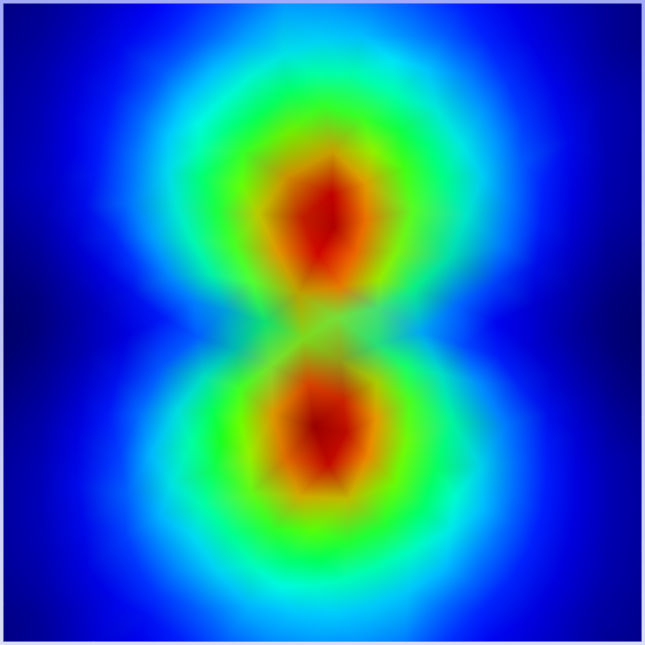}}
   \subfigure{\includegraphics[width=0.23\columnwidth]{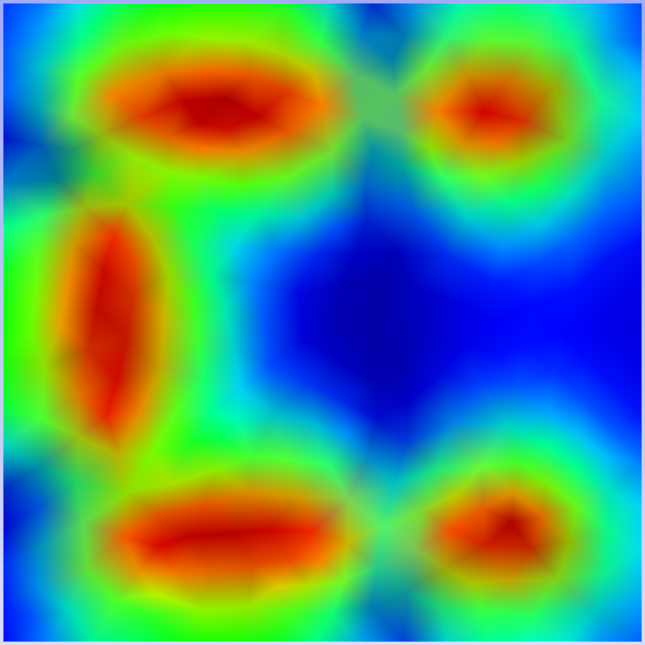}}	
   \subfigure{\includegraphics[width=0.23\columnwidth]{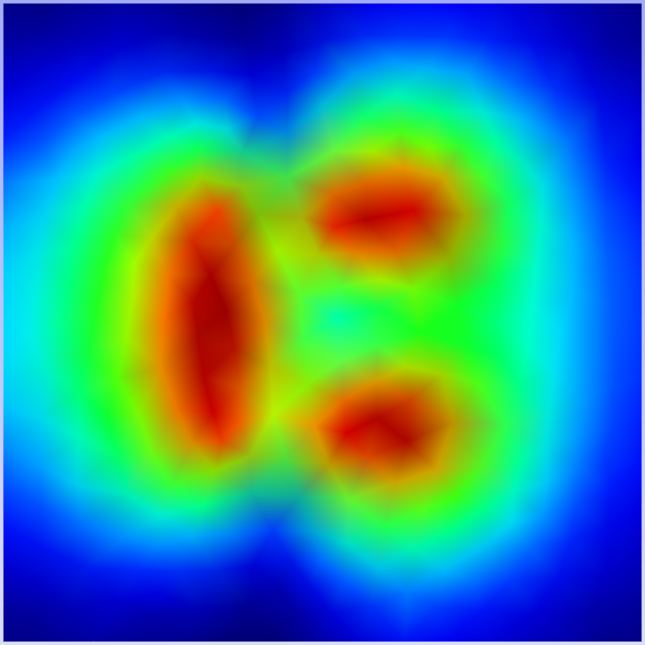}}
   \subfigure{\includegraphics[width=0.23\columnwidth]{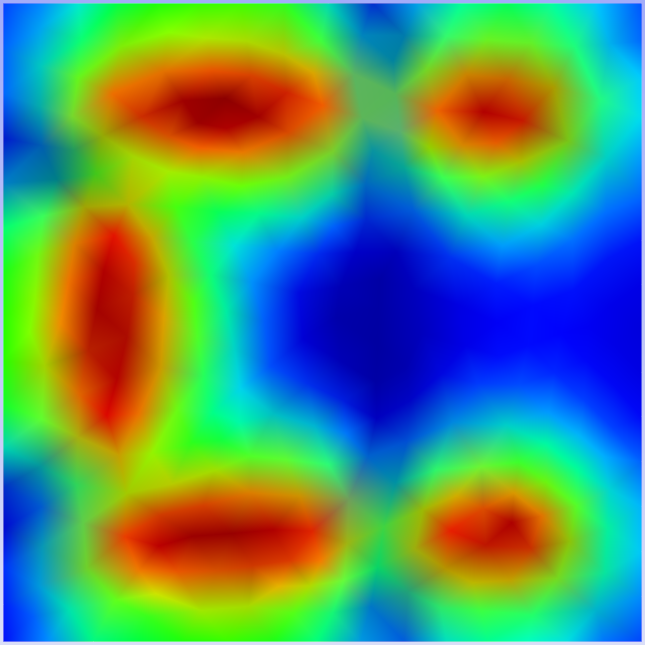}}
  \subfigure{\includegraphics[width=0.23\columnwidth]{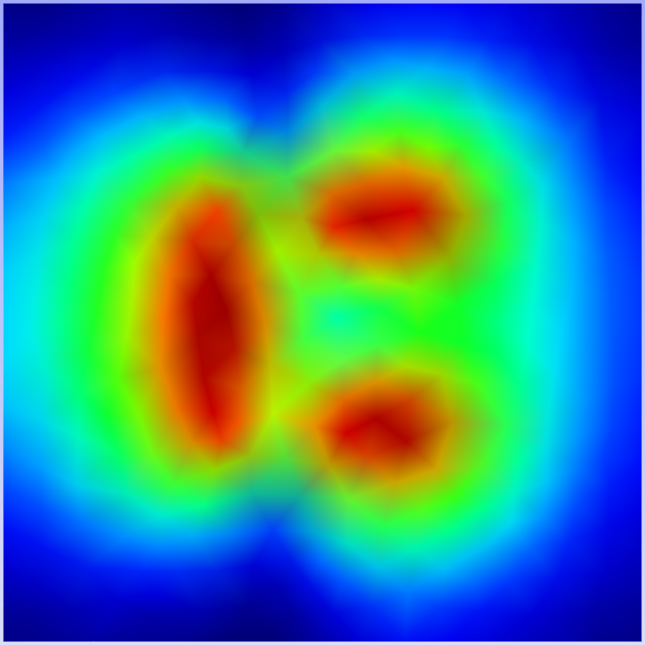}}
  \subfigure{\includegraphics[width=0.23\columnwidth]{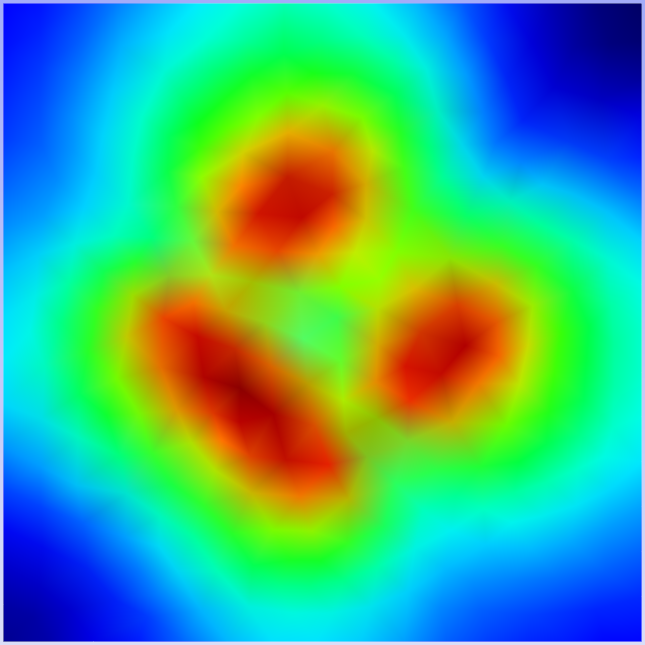}}
   \subfigure{\includegraphics[width=0.23\columnwidth]{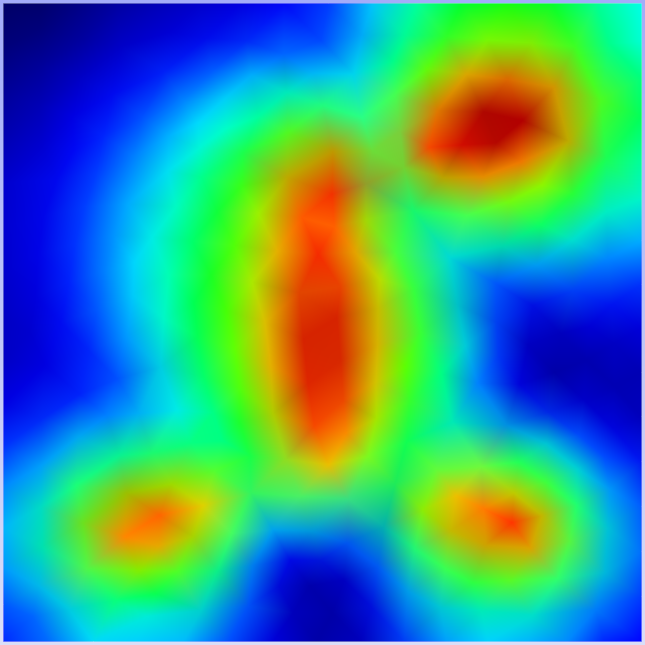}}
   \subfigure{\includegraphics[width=0.23\columnwidth]{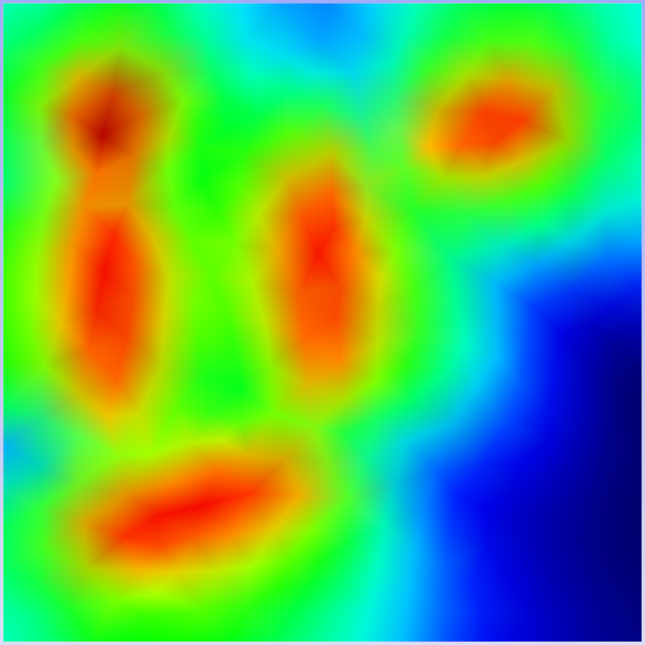}}
   \subfigure{\includegraphics[width=0.23\columnwidth]{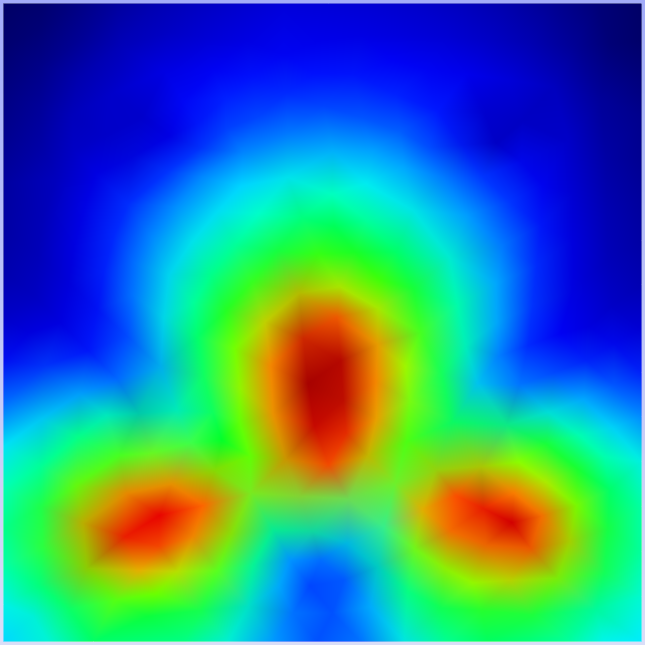}}
   \subfigure{\includegraphics[width=0.23\columnwidth]{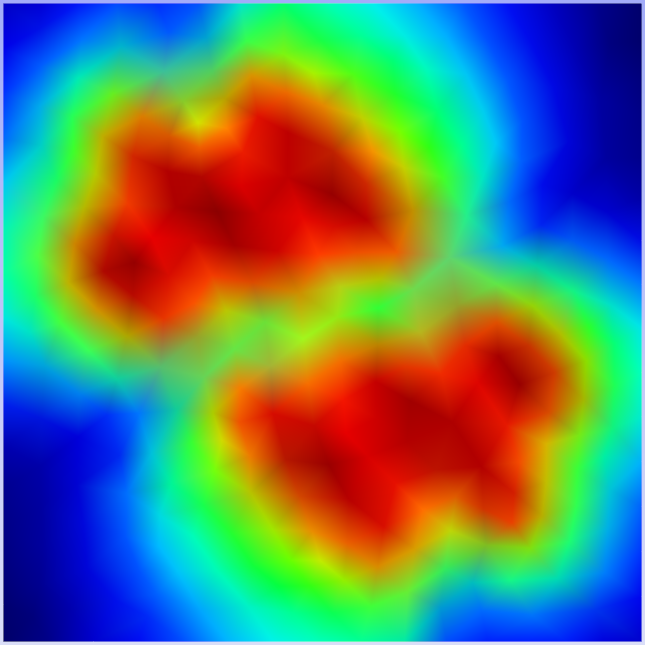}}
   \subfigure{\includegraphics[width=0.23\columnwidth]{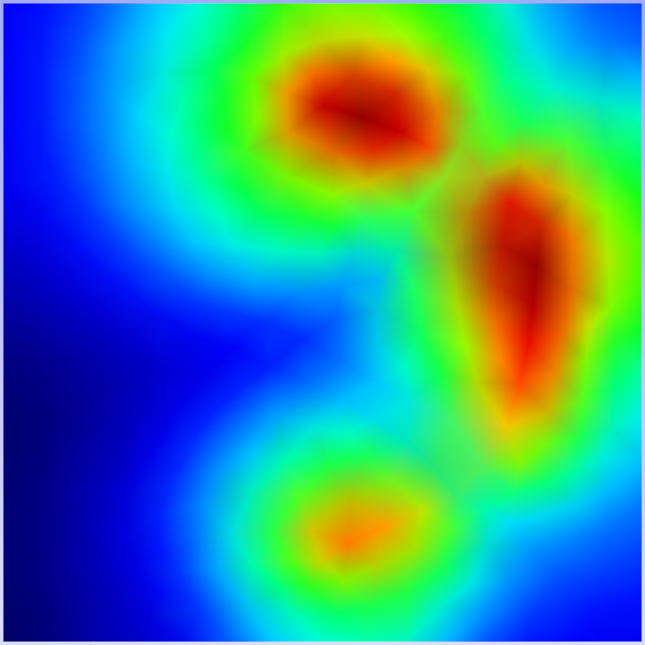}}
   \subfigure{\includegraphics[width=0.23\columnwidth]{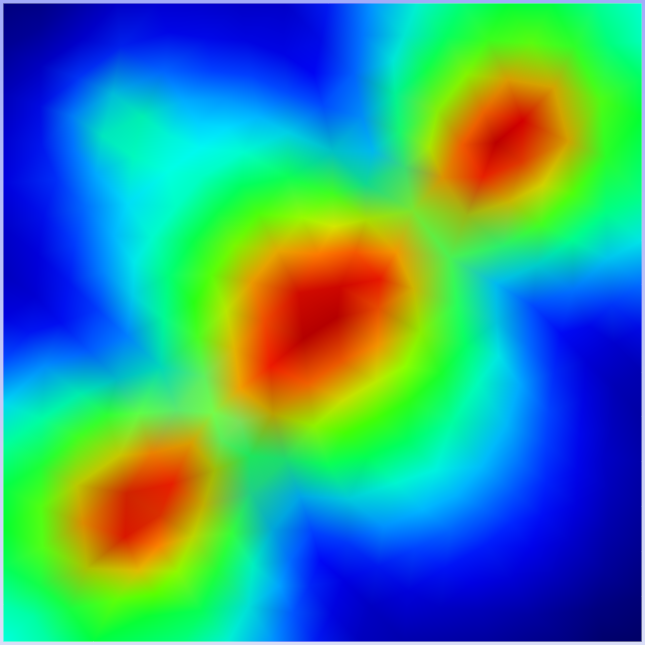}}
   \subfigure{\includegraphics[width=0.23\columnwidth]{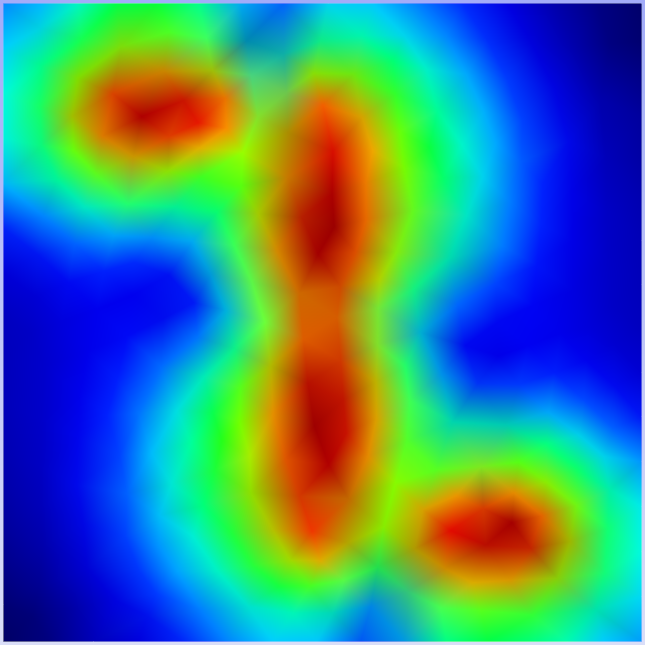}}
   \subfigure{\includegraphics[width=0.23\columnwidth]{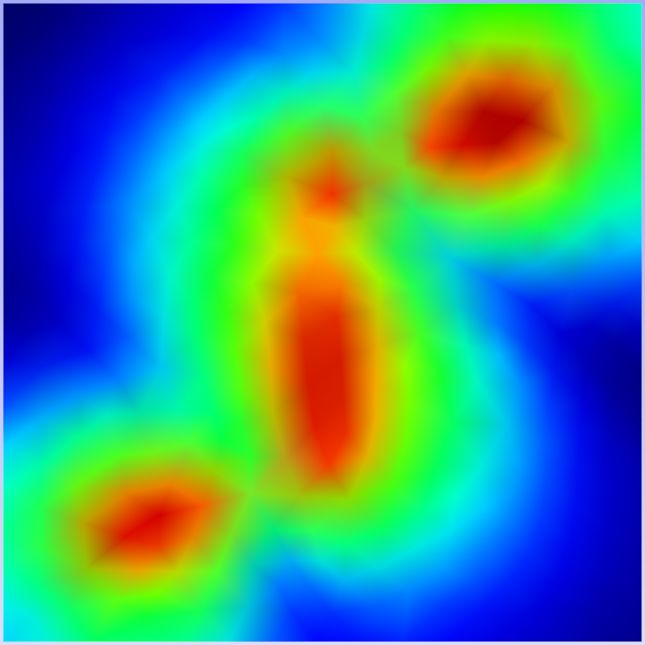}}
   \subfigure{\includegraphics[width=0.23\columnwidth]{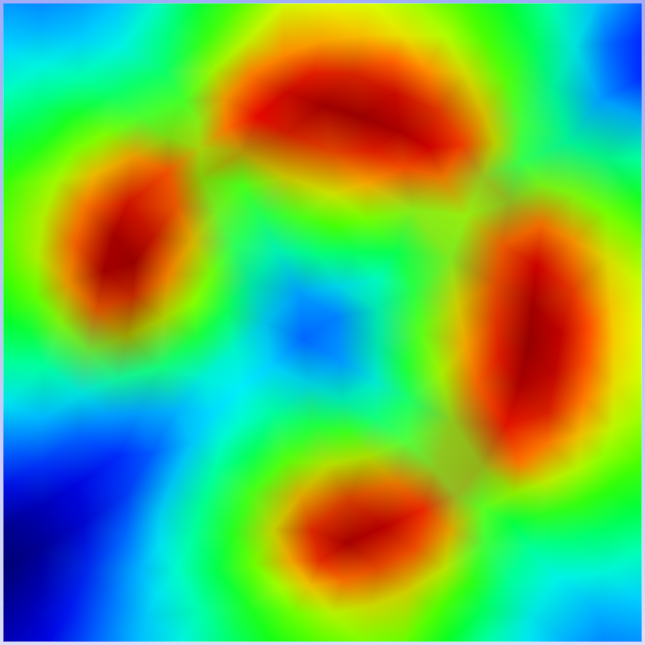}}
   \end{center} 
\caption{Magnitude of the radiated electric field from 32 radiating open wires with different shapes.} 
\label{Fig4} 
\end{figure}

The scan table is considered as an infinite perfect conductor whose potential is set equal to the reference GND. The scanning probes can cover the 2D sampling range of 30\,cm $\times$ 30\,cm. Overall, 64 different shapes of wire straps (32 closed and 32 open endings) are routed parallel to the GND layer at 1cm distance to the table. The conducting wires are excited by a single 1V voltage source randomly placed on the middle, corner, or ending segment of the wires. The wires radius is 1mm and the frequency of voltage sources is set to 1\,GHz for all experiments. The magnitude of the radiated magnetic fields in Fig. \ref{Fig1} and Fig. \ref{Fig2} or the radiated electric fields in Fig. \ref{Fig3} and Fig. \ref{Fig4} are observed on 2\,cm above the table surface in dB scale. Dataset 1 includes 32 bended electric dipoles containing U-form monopoles with straight junctions and the side length of mostly 10\,cm, whereas dataset 0 encompasses 32 polygonal magnetic dipoles with the side length of around 10\,cm. The diversity of database was enhanced by irregularity via including shrunken down variants to the half size and partial rounding or rotating the wires. The method of moments (MoM) is utilised to compute the radiated electromagnetic fields \cite{b0}. The captured scan images in Fig. \ref{Fig1} and Fig. \ref{Fig2} are obtained using equally distanced magnetic field sensors for three orthogonal orientations. Similarly, the Fig. \ref{Fig3} and Fig. \ref{Fig4} are collected by 30$\times$30 arrays of electric field probes in three {\it x, y, z} directions.  

The color images are first converted into grayscale images. The images are resized to 100$\times$100 before being used as the train data, i.e. Gaussian smoothing are performed when downsampling to avoid aliasing artifacts. In contrary to \cite{b1}, no feature extraction method is applied here. In fact, image pixels are selected as features. The samples are labeled either as 0 representing the close or 1 denoting open conductors.

\section{Supervised Learning} \label{sec:results}

Near-field scan images shown in Fig. \ref{Fig1} and Fig. \ref{Fig2} are used as the train set. Leave-one-out (LOO) cross-validation is used to make sure that the predicted results are unbiased to the test sets, i. e. each observation is left out as the validation set, whereas the remaining original samples are used as the training set. The test error is then calculated on the hold out images. The simulation has been run on a 2,3 GHz quad-core Intel processor with 16 GB random access memory under the macOS operating system \cite{b2}. Table~\ref{Table1} demonstrates the harmonic average of precision and recall, the so-called F$_1$ score. Remind that the precision is number of true positive results divided by the number of all positive results returned by the classifier, and recall is the number of true positive results divided by the number of all samples that should have been identified as positive. 

The support vector machine (SVM) is an effective supervised learning method when number of dimensions is greater than the number of samples. The radial basis function (RBF) kernel with coefficient $\gamma=0.001$ is used here.  For the {\it k}-nearest neighbors algorithm, non-parametric ({\it k}-NN) method, $k=3$ is set. The Gaussian process classification (GPC) based on Laplace approximation exhibits the best results. The kernel's hyperparameters are optimized during fitting. Using the decision tree classifier (DTC) no maximum depth is set for the tree to let the DTC expands the nodes splitting until all leaves are pure. F$_1$ score round 0.766 is obtained depending to the integer number given to the seed used by the random state generator. The Gaussian Naive Bayes (GNB) works on electric probe datasets better than the magnetic probe datasets. Using the quadratic discriminant analysis F$_1$ score reaches to 0.75 for the databases gathered with magnetic field probes. Using the nearest centroid classifier F$_1$ score reaches to 0.734 for the databases gathered with electric field probes. Table \ref{Table1} also shows that the stochastic gradient descent (SGD) classifier is inferior to the AdaBoost-SAMME algorithm or random forest (RF) estimator with 100 max features. 

\begin{table}
\caption{F$_1$ score of classifiers trained by the magnetic or electric datasets}
\label{table}
\setlength{\tabcolsep}{3pt}
\hspace{13mm}
\begin{tabular}{|p{50pt}|p{50pt}|p{50pt}|}
\hline
\vspace{0.01mm}
Classifier& 
\vspace{0.01mm}
Magnetic field& 
\vspace{0.01mm}
Electric field\\
\hline
\vspace{0.01mm}
SVM& 
\vspace{0.01mm}
0.797& 
\vspace{0.01mm}
0.828\\
{\it k}-NN& 
0.813& 
0.781\\
GPC& 
0.828& 
0.859\\
GNB& 
0.703& 
0.812\\
MLP& 
0.734& 
0.843\\
SGD& 
0.672& 
0.652\\
AdaBoost& 
0.750& 
0.703\\
DTC& 
0.766& 
0.766\\
RF& 
0.719& 
0.750\\

\hline
\end{tabular}
\label{Table1}
\end{table} 

\section{Conclusion}
Near-field data of a scan table were used as the training set to automatically detect and distinct the radiating open endings from closed routings without using any feature descriptor. The averaged F$_1$ score of almost 90 percent implies the possibility of categorising the type of radiators from their magnitude only emitted near-field pattern. 
The output of this research is a software package well-suited to be retrained based on any measured near-field databank to automate the identification of magnetic-type or electric-type of the radiating coupling sources. 

\section*{Disclaimer}
This study has been preformed solely using the author's own resources in his spare time. This report has not benefited from properties of a particular organisation, and hence, it does not reflect the view of any associations.


\begin{thebibliography}{00} 

\bibitem{b3} D. Deschrijver, F. Vanhee, D. Pissoort, T. Dhaene, ``Automated near-field scanning algorithm for the {EMC} analysis of electronic devices,'' \emph{IEEE Trans. Electromagn. Compat.}, vol. 54, no. 3, pp. 502--510, 2012. 

\bibitem{b0} A. Geranmayeh, R. Moini, S.H. Sadeghi, A. Deihimi, ``A fast wavelet-based moment method for solving thin-wire EFIE,'' \emph{IEEE Trans. Magn.}, vol. 42, no. 4, pp. 575--578, 2006. 

\bibitem{b1} Q. Huang and J. Fan, ``Machine learning based source reconstruction for {RF} desense,'' \emph{IEEE Trans. Electromagn. Compat.}, vol. 60, no. 6, pp. 1640--1647, 2018. 

\bibitem{b2} F. Pedregosa, G. Varoquaux, A. Gramfort, V. Michel, B. Thirion, O. Grisel, {\it et al.}, ``Scikit-learn: Machine Learning in {P}ython,'' \emph{Journal of Machine Learning Research}, vol. 12, pp. 2825--2830, 2011. 

\end{thebibliography}
\end{document}